\documentclass[fleqn,usenatbib]{mnras}

\usepackage{newtxtext,newtxmath}

\usepackage[T1]{fontenc}

\DeclareRobustCommand{\VAN}[3]{#2}
\let\VANthebibliography\thebibliography
\def\thebibliography{\DeclareRobustCommand{\VAN}[3]{##3}\VANthebibliography}

\usepackage{graphicx}	
\usepackage{amsmath}	
\usepackage{chemformula}
\usepackage[separate-uncertainty = true,multi-part-units=single,group-separator={,}, group-digits = integer, group-four-digits = true,per-mode=reciprocal]{siunitx}
\usepackage{multirow}

\newcommand{\matthewred}[0]{\texttt{Eureka!} }
\newcommand{\tspectro}[0]{\texttt{transitspectroscopy} }
\newcommand{\merlinred}[0]{\texttt{Merlin} }

\title[NIRSpec Transmission Spectra of TOI-1685~b]{JWST NIRSpec finds no clear signs of an atmosphere on TOI-1685~b}

\author[Fisher \& Hooton et al.]{
Chloe E.\ Fisher$^{1}$\thanks{Both authors contributed equally}\thanks{E-mail: chloe.fisher@physics.ox.ac.uk}, 
Matthew J.\ Hooton$^{2}$\footnotemark[1]\thanks{Email: mh2143@cam.ac.uk}, 
Amélie Gressier$^{3}$, 
Merlin Zgraggen$^{4}$, 
Meng Tian$^{5}$, \newauthor
Kevin Heng$^{5,6,7,8}$,
Natalie H.\ Allen$^{9}$, 
Richard D.\ Chatterjee$^{10}$, 
Brett M.\ Morris$^{3}$, \newauthor
Nicholas W.\ Borsato$^{11,12,13}$,
Néstor Espinoza$^{3,9}$, 
Daniel Kitzmann$^{14,4}$,
Tobias G.\ Meier$^{1,10}$, \newauthor
Lars A.\ Buchhave$^{15}$,
Adam J.\ Burgasser$^{16}$,
Brice-Olivier Demory$^{4,14,6}$,
Mark Fortune$^{17}$, \newauthor
H.\ Jens Hoeijmakers$^{11}$,
Raphael Luque$^{18,19}$,
Erik A.\ Meier Vald\'es$^{1}$,
João M.\ Mendonça$^{20,21}$, \newauthor
Bibiana\ Prinoth$^{11}$,
Alexander D.\ Rathcke$^{15}$,
and Jake Taylor$^{1}$
\\\\
$^{1}$ Astrophysics, Department of Physics, Parks Road, Oxford, OX1 3RH, UK\\
$^{2}$ Cavendish Laboratory, JJ Thomson Avenue, Cambridge CB3 0HE, UK\\
$^{3}$ Space Telescope Science Institute, 3700 San Martin Drive, Baltimore, MD 21218, USA\\
$^{4}$ Center for Space and Habitability, University of Bern, Gesellschaftsstrasse 6, 3012 Bern, Switzerland\\
$^{5}$ Faculty of Physics, Ludwig Maximilian University, Scheinerstrasse 1, D-81679, Munich, Bavaria, Germany.\\
$^{6}$ ARTORG Center for Biomedical Engineering Research, University of Bern, Murtenstrasse 50, CH-3008, Bern, Switzerland\\
$^{7}$ University College London, Department of Physics \& Astronomy, Gower St, London, WC1E 6BT, United Kingdom \\
$^{8}$ Astronomy \& Astrophysics Group, Department of Physics, University of Warwick, Coventry CV4 7AL, United Kingdom\\
$^{9}$ William H. Miller III Department of Physics and Astronomy, Johns Hopkins University, Baltimore, MD 21218, USA \\
$^{10}$ Atmospheric, Oceanic and Planetary Physics, Department of Physics, University of Oxford, Oxford, United Kingdom \\
$^{11}$ Lund Observatory, Division of Astrophysics, Department of Physics, Lund University, Box 118, 221 00 Lund, Sweden\\
$^{12}$ School of Mathematical and Physical Sciences, Macquarie University, Sydney, NSW 2109, Australia\\
$^{13}$ Astrophysics and Space Technologies Research Centre, Macquarie University, Sydney, NSW 2109, Australia \\
$^{14}$ Weltraumforschung und Planetologie, Physikalisches Institut, University of Bern, Gesellschaftsstrasse 6, 3012 Bern, Switzerland\\
$^{15}$ DTU Space, Technical University of Denmark, Elektrovej 328, DK-2800 Kgs. Lyngby, Denmark\\
$^{16}$ Department of Astronomy \& Astrophysics, UC San Diego, La Jolla, CA 92093, USA \\ 
$^{17}$ School of Physics, Trinity College Dublin, University of Dublin, Dublin 2, Ireland \\
$^{18}$ Department of Astronomy \& Astrophysics, University of Chicago, Chicago, IL 60637, USA\\
$^{19}$ NHFP Sagan Fellow\\
$^{20}$ School of Physics and Astronomy, University of Southampton, Highfield, Southampton SO17 1BJ, UK \\
$^{21}$ School of Ocean and Earth Science, University of Southampton, Southampton, SO14 3ZH, UK \\
}

\date{Accepted XXX. Received YYY; in original form ZZZ}

\pubyear{\the\year{}}

\begin{document}
\label{firstpage}
\pagerange{\pageref{firstpage}--\pageref{lastpage}}
\maketitle

\begin{abstract}
Determining the prevalence of atmospheres on terrestrial planets is a core objective in exoplanetary science. While M dwarf systems offer a promising opportunity, conclusive observations of terrestrial atmospheres have remained elusive, with many yielding flat transmission spectra. We observe four transits of the hot terrestrial planet TOI-1685~b using JWST's NIRSpec G395H instrument. Combining this with the transit from the previously-observed phase curve of the planet with the same instrument, we perform a detailed analysis to determine the possibility of an atmosphere on TOI-1685~b. From our retrievals, the Bayesian evidence favours a simple flat line model, indicating no evidence for an atmosphere on TOI-1685~b, in line with results from the phase curve analysis. Our results show that hydrogen-dominated atmospheres can be confidently ruled out. For heavier, secondary atmospheres we find a lower limit on the mean molecular weight of $\gtrsim10$, at a significance of $\sim5\sigma$. Pure \ch{CO2}, \ch{SO2}, \ch{H2O}, and \ch{CH4} atmospheres, or a mixed secondary atmosphere (\ch{CO}+\ch{CO2}+\ch{SO2}) could explain the data ($\Delta\ln Z<3$). However, pure \ch{CH4} atmospheres may be physically unlikely, and the pure \ch{H2O} and \ch{CO2} cases require a high-altitude cloud, which could also be interpreted as a thin cloud-free atmosphere. We discuss the theoretical possibility for different types of atmosphere on this planet, and consider the effects of atmospheric escape and stellar activity on the system. Though we find that TOI-1685~b is likely a bare rock, this study also highlights the challenges of detecting secondary atmospheres on rocky planets with JWST.  
\end{abstract}




\section{Introduction}
\label{sec:intro}

\subsection{Terrestrial Planets with JWST}
\label{sec:superearths_jwst}

Terrestrial exoplanets are those with a bulk composition likely dominated by rock or iron, and that possess a solid or liquid surface. Super-Earths, with a radius of 1-2$\times$ that of Earth, and smaller exoplanets are expected to be terrestrial. These planets have been challenging to observe, due to their smaller radius and lower atmospheric scale height compared to gaseous exoplanets. Previous studies of rocky exoplanets from HST and Spitzer have been restricted to ruling out mostly cloud-free, hydrogen-dominated atmospheres. The launch of JWST represents a new era of terrestrial planet observations, with the ability to observe spectra at a much higher precision and across a wider wavelength range. Cycles 1-4 include observations of over 50 different rocky exoplanets, spanning a variety of radii, masses, and equilibrium temperatures. These observations will determine the prevalence of atmospheres around terrestrial planets, particularly those orbiting M dwarf stars \citep[see, e.g.,][for a review of these JWST programs and observations]{espinoza25}.

\subsubsection{Observations in Transmission}

To date, JWST transmission spectroscopy of terrestrial planets has predominantly yielded featureless spectra, likely due to their higher mean molecular weights and/or cloudy atmospheres \citep{lustig-yaeger23,kirk24,scarsdale24,alderson24,alderson25,alam25,taylor25,piaulet-ghorayeb25}. However, several terrestrial planets have displayed intriguing signals in their transmission spectra, though the interpretation remains a challenge. In particular, contamination from stellar activity can have a big impact on transmission spectra, often dominating the spectrum with distinct stellar features, particularly in the case of very active host stars such as Trappist-1 \citep{lim23,radica24b,piaulet-ghorayeb25}. In other cases, stellar activity has led to several spectra of super-Earths containing features that could be attributed to either the star or the planet \citep{may23,rackham23,moran23a}. Further observations have been able to confirm the origin of these features as stellar \citep{xue24,weinermansfield24,bennett25}. This demonstrates additional challenges for measuring the atmospheres on terrestrial exoplanets, particularly those orbiting active stars. The results advocate for a wider wavelength coverage, and the necessity for multiple consistent observations before a confident detection of an atmosphere can be claimed. 

Tentative signs of an atmosphere have been detected from the transmission spectra of a few super-Earths. Sulfur species have been proposed to explain the observations of L 98-59 d \citep{banerjee24,gressier24} and L 98-59 b \citep{bello-arufe25}. This could be due to high levels of volcanism on these planets. However, further observations will be necessary to confirm the potential presence of these molecules. A possible \ch{N2}-dominated atmosphere has been suggested for the planet LHS 1140~b \citep{damiano24,cadieux24}, which, based on its density, could either be a mini-Neptune with a small hydrogen envelope or a super-Earth with a large fraction of water by mass -- a so-called ``water world" \citep{cadieux24a}. The lack of evidence for a hydrogen-dominated atmosphere supports the water world scenario, but follow-up observations should provide a more definitive conclusion.

\subsubsection{Observations in Emission}

The atmospheres of small exoplanets are likely to be secondary -- i.e. sourced through geochemical outgassing, as opposed to through accretion during the planet's formation. Hot terrestrial planets are prime targets for studying such atmospheres, as high surface temperatures can drive outgassing of \ch{CO2} or silicate species \citep[e.g.,][]{gaillard14,zilinskas23}. These atmospheres have high mean molecular weights and low scale heights, making them difficult to detect in transit, but their high temperatures favour secondary eclipse observations. As a result, several JWST programs, including large surveys, are targeting hot rocky planets in secondary eclipse. One such survey is the Hot Rocks program (JWST GO 3730; PI Diamond-Lowe), observing secondary eclipses of nine irradiated terrestrial planets with MIRI's \SI{15}{\micron} filter. So far, one planet shows tentative evidence for an atmosphere \citep{august25}, whilst three others are consistent with a bare rock \citep{meiervaldes25,fortune25,allen25}. A similar, larger survey is now being conducted as part of the Rocky Worlds DDT Program. Secondary eclipse measurements with MIRI have also supported the bare rock scenario for Trappist-1~b \citep{greene23,ducrot25}, and disfavoured a thick \ch{CO2} atmosphere on Trappist-1~c \citep{zieba23}. Emission spectra, particularly of very hot terrestrial planets that may contain magma oceans on their surfaces, can also constrain super-Earth atmospheres, and some tentative evidence has been found for a \ch{CO}- or \ch{CO2}-rich atmosphere on 55~Cancri~e \citep{hu24}. Although these ultra hot objects are distinctly different in nature from the cooler terrestrial planets previously discussed, the observation of their secondary atmospheres would provide clues into the evolution of the disequilibrium chemistry and the increased heat redistribution likely to be present in the atmospheres of their cooler counterparts.

\subsection{The TOI-1685 System}

The super-Earth TOI-1685~b orbits an M dwarf host \citep[M2.5V, $M_\star=0.454\pm0.018M_\odot$, $R_\star=0.4555\pm0.0128R_\odot$,][]{terrien15,burt24} with a period of 0.7 days, giving it an equilibrium temperature of $1062\pm27$K. Originally the subject of two simultaneous discovery papers with conflicting stellar and planet parameters \citep{bluhm21,hirano21}, TOI-1685~b had the potential to be one of the most optimal targets for atmospheric detection with JWST. Using data from TESS and Carmenes, \cite{bluhm21} determined the planet radius and mass to be $1.70\pm0.07~R_\oplus$ and $3.78\pm0.63~M_\oplus$, respectively, resulting in a bulk density of $4.21^{+0.95}_{-0.82}~$\SI{}{\g\per\cm\cubed}. \cite{hirano21} constrained the planet radius and mass to be $1.459\pm0.065~R_\oplus$ and $3.43\pm0.93~M_\oplus$, respectively. This puts the planet density at around $6.1^{+1.9}_{-1.7}~$\SI{}{\g\per\cm\cubed}. \cite{luque22} combined the data from both papers, and found planet parameters in strong agreement with \cite{bluhm21}. However, they adopted the stellar parameters from \cite{bluhm21} in their analysis, which could account for the agreement. A recent study by \cite{burt24} performed the most comprehensive analysis so far, combining the previous data sets with new TESS sectors, as well as ground-based RVs from MAROON-X, and re-determining the stellar parameters from scratch. They found a much closer agreement with the parameters from \cite{hirano21}, measuring the planet radius and mass to be $1.468^{+0.050}_{-0.051}~R_\oplus$ and $3.03^{+0.33}_{-0.32}~M_\oplus$, respectively. These parameters indicate an Earth-like bulk density ($5.3\pm0.8~$\SI{}{\g\per\cm\cubed}). 

\begin{figure}
    \centering
    \includegraphics[width=0.5\textwidth]{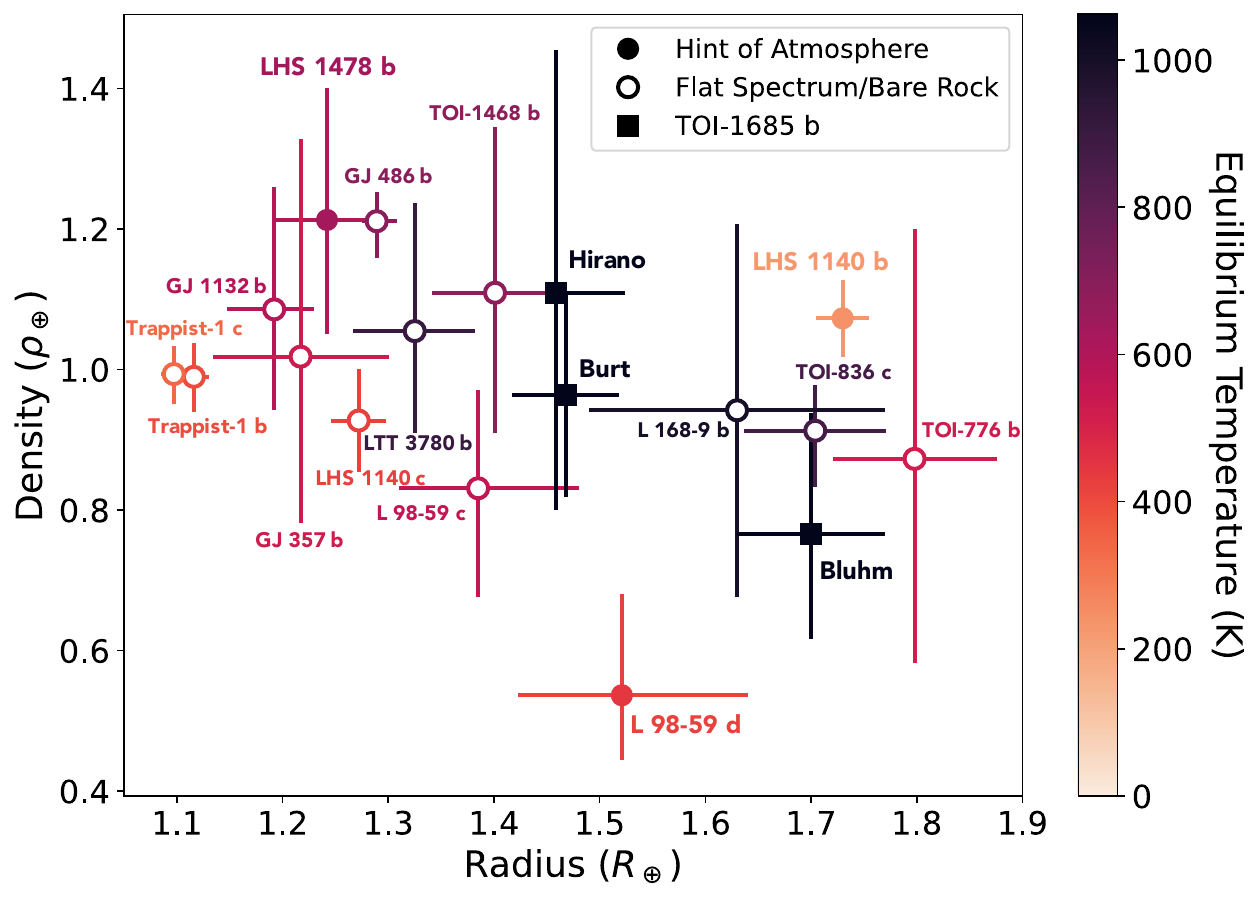}
    \caption{Density versus radius values for super-Earths observed with JWST, coloured by their equilibrium temperature. Filled circles represent planets with hints of an atmosphere, whilst empty circles represent flat transmission spectra, or secondary eclipse measurements indicating bare rocks. The squares show the measurements for TOI-1685~b from various papers.}
    \label{fig:density-radius}
\end{figure}

Figure \ref{fig:density-radius} shows the density and radius values of super-Earths observed with JWST, discussed in Section \ref{sec:superearths_jwst}. The values for TOI-1685~b from \cite{hirano21}, \cite{bluhm21}, and \cite{burt24} are also shown, indicating the wide variability of reported values from the initial discovery papers and follow-up studies. The density from \cite{bluhm21} is too low to be consistent with a purely rock planet, implying it could either have a substantial atmosphere or a significant (50\%) water fraction. The latter would class it as a planet with a high fraction of water in its interior, and a volatile-rich envelope. However, the higher density from \cite{hirano21} and \cite{burt24} is inconsistent with these scenarios, instead suggesting TOI-1685~b is a rocky planet with negligible volatiles. 

Finally, TOI-1685~b was recently observed by CHEOPS and analysed as part of a study considering so-called Hot Water World triangles \citep{egger25}. This refers to the region of mass-radius parameter space where the density indicates that the planet must contain volatiles in its envelope, though these envelopes are not necessarily water-dominated. This model is specific to close-in (i.e. hot) planets, with orbital periods $\lesssim10$ days, where thin H/He envelopes would be unstable due to the stellar irradiation. The CHEOPS observations give a planet radius of $1.421\pm0.060~R_\oplus$, consistent with the values from \cite{hirano21} and \cite{burt24}. This places the planet outside the Hot Water World triangle, in agreement with the conclusions from \cite{burt24}.

Based on the lower density predicted for TOI-1685~b in \cite{bluhm21}, three separate JWST proposals were approved in cycle 2 to observe this planet. The first observed a phase curve using NIRSpec G395H (JWST GO 3263; PI Luque), the second observed two transits using NIRISS SOSS as part of a larger survey of potential water worlds (JWST GO 4098; PI Benneke), and the third observed four transits using NIRSpec G395H (JWST GO 4195; PI Fisher), presented in this study. The results of the phase curve were recently published in \cite{luque25}, who reported that the emission spectrum and phase curve amplitude and offset are consistent with a blackbody with no heat redistribution and a low albedo. They also analysed the transmission spectrum, and were able to rule out cloud-free \ch{H2}-dominated atmospheres. However the single transit does not have the sensitivity to rule out secondary atmospheres, particularly in the presence of clouds. In this study, we present our results from our four transits with NIRSpec, and include a re-analysis of the additional transit from the phase curve, thanks to the same instrument set-up. 

\subsection{Layout of the study}

In Section \ref{sec:observations} we describe the five NIRSpec observations of TOI-1685~b that form the basis of the study. In Section \ref{sec:data_reductions} we present our approach to reducing the data and fitting for the transits using three independent reduction codes. In Section \ref{sec:analysis} we describe our analysis methods and present the results from Gaussian tests and our atmospheric modeling. In Section \ref{sec:discussion} we discuss the implications of our results and compare to other recent studies of terrestrial planets. Finally, in Section \ref{sec:conclusions} we present our conclusions. 

\section{Observations}
\label{sec:observations}

We observed transits of TOI-1685~b four times as part of JWST GO 4195 using JWST with the NIRSpec instrument in its Bright Object Time Series (BOTS) mode. The observations were conducted using the F290LP filter and G395H grism providing wavelength coverage of 2.87 - 3.72 \unit{\micro\meter} on the NRS1 detector and 3.82 - 5.17 \unit{\micro\meter} on the NRS2 detector. For all four visits, we used the SUB2048 subarray with 16 groups per integration and 1032 integrations during the single exposure, yielding an integration time of 15.35 s and a visit duration of 264 minutes. Given the 53 minute duration of the TOI-1685~b transit, this meant that each visit had roughly four transit durations worth of observations from which to accurately measure the out-of-transit flux from the host star and search for the presence of other time-correlated noise. The start times of each visit were 2024 February 25 at 08:04 UTC, 2024 February 29 at 08:20 UTC, 2024 March 02 at 08:31 UTC, and 2024 September 01 at 00:48 UTC. While the fourth visit was originally scheduled shortly after the other three, a failed observation led to its delay for a few months until the target became visible with JWST again. 

As described in Section \ref{sec:intro}, TOI-1685 was also observed on 2024 February 15 at 07:45 UTC using NIRSpec/BOTS with the G395H grism as part of JWST GO 3263. While the strategy of this visit was almost identical to that deployed in our visits, the main difference was that the visit duration of 19.0 hr covered just over a full orbit of TOI-1685~b. While we considered a full joint analysis of the full-phase curve unlikely to improve upon the work presented in \citet{luque25}, we welcomed the addition of the transit contained within these observations to boost the SNR on the transmission spectrum and to help look for evidence of time-evolution in the data. To make this visit directly comparable to the visits from our programme, we trimmed the visit to be the same length as ours, centered on the transit.

\section{Data Reduction \& Light Curve Fitting}
\label{sec:data_reductions}

To check how our extracted transmission spectrum varies due to differences in reduction and light curve fitting strategies, we independently performed three different methodologies, which we describe in detail in the following 3 sections.

\subsection{Methodology 1: \matthewred}
\label{sec:eureka}

\begin{figure*}
    \includegraphics[]{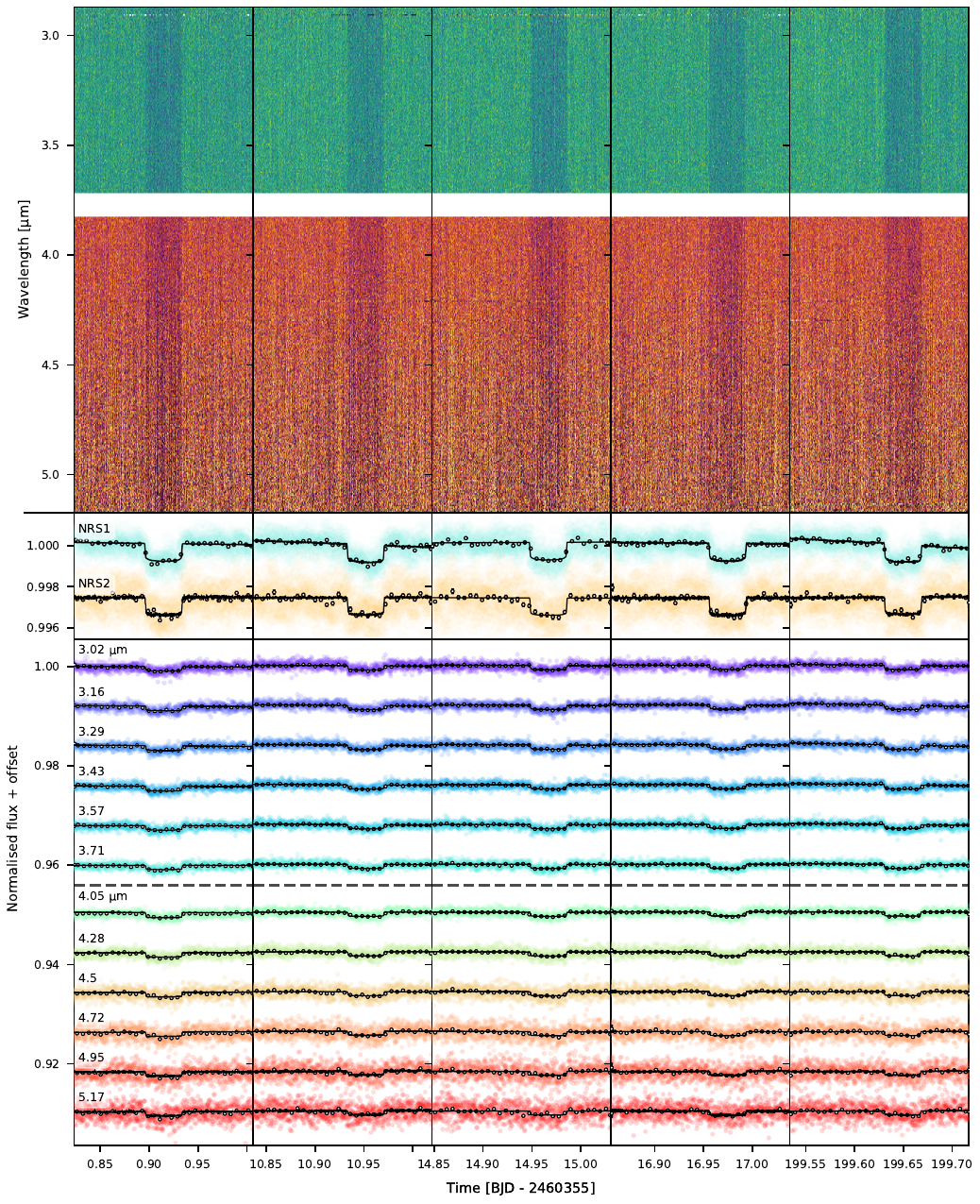}
    \caption{The light curves from the NIRSpec G395H observations of TOI-1685, with the five visits shown chronologically from left to right. Top panel: raw, undetrended light curves at wavelength resolution of 6.5 nm, which was used in our adopted fit. NRS1 is shown in blue colours and NRS2 is shown in red colours, with the darkening in the centre of each panel corresponding to the flux drop during the transit of TOI-1685~b. Middle panel: Raw, undetrended NRS1 and NRS2 white light curves for each visit. Unbinned data are shown in colour, with data in bins of 9 minutes shown in empty black markers, with error bars generally too small to be visible. Bottom panel: Light curves in 6 wavelength bins across each detector in the same structure as the middle panel, with the detectors separated by a dashed line.}
    \label{fig:lcs}
\end{figure*}

To reduce the raw time-series spectra from the observation, we used version 1.2 of the \matthewred pipeline \citep{bell22}. For Stage 1, we used the standard \texttt{jwst} pipeline output (rateints files) which converts ramps to slopes via ramp fitting. The \texttt{clean\_flicker\_noise} step, which is the \texttt{jwst} pipeline's standard correction of 1/f noise, was not applied, and the jump detection threshold was left at the default value. We began with the rateints files that had been converted from ramps to slopes using Stage 1 of the \texttt{jwst} pipeline. Stage 2 follows the standard \texttt{jwst} pipeline but skips the \texttt{photom} step, which is unnecessary for measuring transit depths. In stage 3, the frames were trimmed outside of x limits of 487 and 2044 for NRS1, and 6 and 2044 and y limits of 0 and 32 for NRS2, the curvature exhibited by the trace for G395H observations was corrected, background was subtracted column by column about a half width of 10 pixels to correct for the 1/f noise. Optimal extraction was used to extract the trace, with its position determined by modelling it as a Gaussian within a half width of 4 pixels. In stage 4, the data were binned in wavelength and light curves for each bin were outputted to an average of 6.5 nm per bin, with a total of 329 bins across both detectors. We manually masked 3 columns in NRS1 and 2 in NRS2 where bad pixels in the aperture resulted in erroneous flux values. To eliminate outliers from each light curve, we clipped points that were over 4$\sigma$ from a rolling median with a width of 10 points. These are shown for all five visits in the top row of Figure \ref{fig:lcs}.

To measure how the absorption at the terminator of TOI-1685~b varies as a function of wavelength across the NIR, we fitted the NIRSpec light curves with models of the planetary transit, other signals arising from the star and instrument. To do this, we used the \texttt{juliet} package, which allows the simultaneous fitting of transiting planets, Gaussian Processes \citep[GPs;][]{gibson12a,gibson14} and other linear basis vectors to account for systematic trends in the light curves.

For the purposes of this model, we assume that the flux that we receive from the star is constant, apart from the drops in flux when planet b is transiting.  We modelled the transit according to \citet{mandel02}, parameterised by the orbital period $P$, the time of inferior conjunction $t_\mathrm{0}$, the ratio of the planetary and stellar radii $R_\mathrm{p}/R_\mathrm{\star}$, the stellar density $\rho_\mathrm{\star}$, the impact parameter $b$, and quadratic limb darkening parameters $u_\mathrm{1}$ and $u_\mathrm{2}$. Assuming a fully circularised orbit at this small separation to its host star, we fixed the eccentricity to 0. 

In addition to fitting the white noise $\sigma_\mathrm{w}$ in the light curves, we considered three sources of time-correlated noise that arise in the instrumentation. The NRS1 detector of NIRSpec exhibits a wavelength-dependent negative slope with time in its time series observations \citep[see e.g.][]{jwsttransitingexoplanetcommunityearlyreleasescienceteam23,alderson23}. We also observed correlations between the full-width half maximum of the trace and light curves from both NRS1 and NRS2. We included linear basis vectors of each of these in the baseline model of the light curves at the white light and spectral level.

Finally, in the 01 Sept visit, quasi-sinusoidal variations of period of 40 minutes and amplitude of roughly 200 ppm were present in both the NRS1 and NRS2 white light curves, and constituted the only strong periodicity that was present in the light curves of any of the five visits. Oscillations may show photometric variations on several timescales due to active regions and rotation, or granulation. We don't expect granulation correlations with timescales near 40 minutes for M3 dwarfs, and the stellar rotation period near $P_{\rm rot} \sim 19$ days is far too long to be implicated. We discuss the stellar rotation and activity in more detail in Section~\ref{sec:activity}. While we are not specifically aware of other examples of similar noise in NIRSpec observations, we consider instrumental noise to be the most likely provenance of this signal. To model this, we fitted a GP as a function of time to the 01 Sept visit light curves. Our kernel of choice was a damped simple harmonic oscillator implemented by \texttt{celerite2}\footnote{\url{https://github.com/exoplanet-dev/celerite2}} package \citep{foreman-mackey17a,foreman-mackey18}, with  hyperparameters of angular frequency $\omega_\mathrm{0}$, amplitude $S_\mathrm{0}$ and quality factor $Q$, the last of these which we fixed to $1/\sqrt{2}$, corresponding to critical damping \citep{gordon20}.

To derive the joint posterior distribution across all of the variables, we used the Dynamic Nested Sampler \citep[see][]{skilling04,skilling06,higson19} in the \texttt{dynesty} package \citep{speagle20}, allowing robust model comparison using its estimates of the Bayesian evidence $\mathcal{Z}$. We use 300 live points and a convergence criterion of $\delta \ln{\mathcal{Z}}<0.01$.

Our starting point was to fit all of the white light curves. In addition to the NRS1 and NRS2 white light curves for each of the five NIRSpec visits, we also included the sector 19 and sector 59 TESS light curves and the 18 ground-based light curves that are presented in \citet[][shown in its Fig 3]{burt24} in the fit. This leads to a more accurate prediction of the timing of each NIRSpec visit for the spectral fits. All of the transit parameters were allowed to vary within wide uniform priors, except for $\rho_\mathrm{\star}$ which had a Gaussian prior of 6.77 $\pm$ 0.63 \unit{g/cm^3} following the stellar modelling in \citet{burt24}. The values of the transit parameters fixed in the spectral light curve fits are shown in Table \ref{tab:transit_params}. The results of this white light curve fit were also used by the spectral light curve fits of the other two methodologies.

\begin{table}
    \centering
    \begin{tabular}{lc}
        \hline
        Parameter & Value \\
        \hline
        $P$ (d) & $0.66913856\pm 0.00000014$ \\
        $t_\mathrm{0}$ (BJD) & $2459593.76594\pm 0.00015$ \\
        $b$ & $0.401_{-0.030}^{+0.031}$ \\
        $\rho_\mathrm{\star}$ (g cm$^{-3}$) & $6.39_{-0.28}^{+0.22}$ \\
        \hline
    \end{tabular}
    \caption{The values and uncertainties of the transit parameters sampled in the white light curve fit. The respective parameters in the spectral fits were fixed to these values.}
    \label{tab:transit_params}
\end{table}


We then fitted all of the spectrally resolved light curves. We fixed the values of $P$, $t_\mathrm{0}$, $\rho_\mathrm{\star}$ and $b$ to the median values obtained in the white light curve fit. $R_\mathrm{p}/R_\mathrm{\star}$ was allowed to vary uniformly between 0 and 0.1. $u_\mathrm{1}$ and $u_\mathrm{2}$ had Gaussian priors placed on them, with mean values determined using values from \citet{burt24} and the MPS-ATLAS-1 grid of stellar models \citep{kostogryz22}, implemented by the \texttt{ExoTiC-LD} package \citep{grant24}. The standard deviation of 0.1 placed on both coefficients reflected both the desire to build an arbitrarily small uncertainty arising our knowledge of the limb darkening where weak constraints are provided by the light curves at the spectral level, along with the disagreements between the values of $u_\mathrm{1}$ and $u_\mathrm{2}$ predicted by different models.

Due to the effect the evolving stellar inhomogeneities can have on JWST transmission spectra of M dwarfs \citep[e.g.][]{lim23}, we performed fits treating each visit individually, as well as fitting all of the visits jointly with shared transit parameters. However, we did not find any major differences between the two methods. Although we fitted each of the 329 light curves across the full range of wavelengths separately, the bottom row of Figures \ref{fig:lcs} and \ref{fig:lcs_detrend} shows the light curves and best fitting transit models in 12 bins for the purpose of visualisation.

\subsubsection{Fitting strategy variations}

\begin{figure}
    \includegraphics{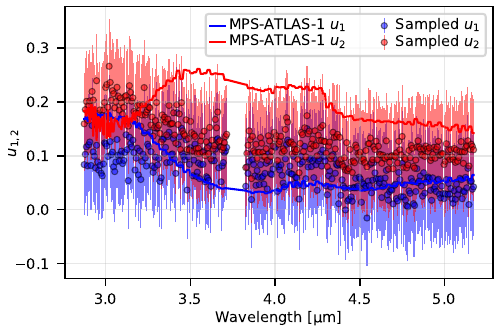}
    \caption{Comparison between the derived MPS-ATLAS-1 values of $u_\mathrm{1}$ and $u_\mathrm{2}$ and those sampled in the combined fit. The values were sampled under a Gaussian prior with a mean of the corresponding MPS-ATLAS-1 value and a standard deviation of 0.1}
    \label{fig:LDs}
\end{figure}

We explored the effect of using an additional two different limb darkening priors. For the first run, we allowed $u_\mathrm{1}$ and $u_\mathrm{2}$ both to vary uniformly within wide limits of -3 and 3. In this case, the relatively low SNR of the transit in the spectrally resolved light curves meant that the data did not constrain $u_\mathrm{1}$ and $u_\mathrm{2}$ particularly well, leading to the sampling of unphysical stellar profiles and a knock on effect on the sampling of other transit parameters. For the second run we fixed $u\mathrm{1}$ and $u\mathrm{2}$ to their MPS-ATLAS-1 values \citep{kostogryz22}, finding that they resulted in a spectrum well within 1 sigma at all wavelengths of the spectrum of the run with Gaussian priors. We opted to stick with the Gaussian priors, which build some uncertainty associated with the true values of $u_\mathrm{1}$ and $u_\mathrm{2}$ into the sampling. The MPS-ATLAS-1 values and the sampled values in the combined fit of $u_\mathrm{1}$ and $u_\mathrm{2}$ are shown in Figure \ref{fig:LDs}.

We also explored the effect of different wavelength bin sizes on the measured transmission spectrum. In addition to the 6.5 nm bins (an average of 10 pixels) described above, we also tested 65 nm bins and 0.65nm bins (100 pixel bins and native resolution). The 100 pixel spectrum was in very good agreement with the 10 pixel spectrum. The native resolution spectrum yielded the same features that were present in the 10 pixel spectrum, but translated downwards by an average of 51 ppm. We treated this as evidence that the limited SNR on the transit at native resolution led to the priors having a greater effect on the fit than when binned. We opted for the 10 pixel binning as a middle-ground between the two options.

We also explored the effect of applying group-level background subtraction in Stage 1 before the ramps were fitted, keeping all other aspects of the reduction and fitting the same. The mean difference in the fitted transit depth across all wavelength bins was 0.077 ppm, with a maximum difference of 0.6 ppm, all within the 1$\sigma$ uncertainties in each case. We judged this aspect to have an insignificant effect on the spectra, so neglected group-level background subtraction for our final fits.

Finally, we explored the effect on the light curve fitting and the resulting transmission spectrum of using GPs in various ways. The visible quasi-sinusoidal signal visible by eye in the 01 Sept visit meant that our minimum requirement for this process was a decent modelling of this noise.  When fitting the visits individually, we found that while the use of this GP had very little effect on the transmission spectra of the first four visits, the effect on the 01 Sept visit was to favour marginally deeper transit depths in 90\% of wavelength bins, albeit in all cases in agreement within 1$\sigma$. We treated this as evidence of a slight bias towards smaller transit depths if this signal was not accounted for. We also tested the effect of using a single GP to fit all of the visits at once, but this led to a less successful fit of the noise in the 01 Sept visit. In the end, we settled on a strategy of only using a GP to fit this visit. This reflects what we judge to be fundamentally different noise properties in the final visit, likely arising due to the 6 months separating it from the previous 4. 

\subsubsection{Individual and Average Spectra}

\begin{figure*}
    \centering
    \includegraphics[width=\textwidth]{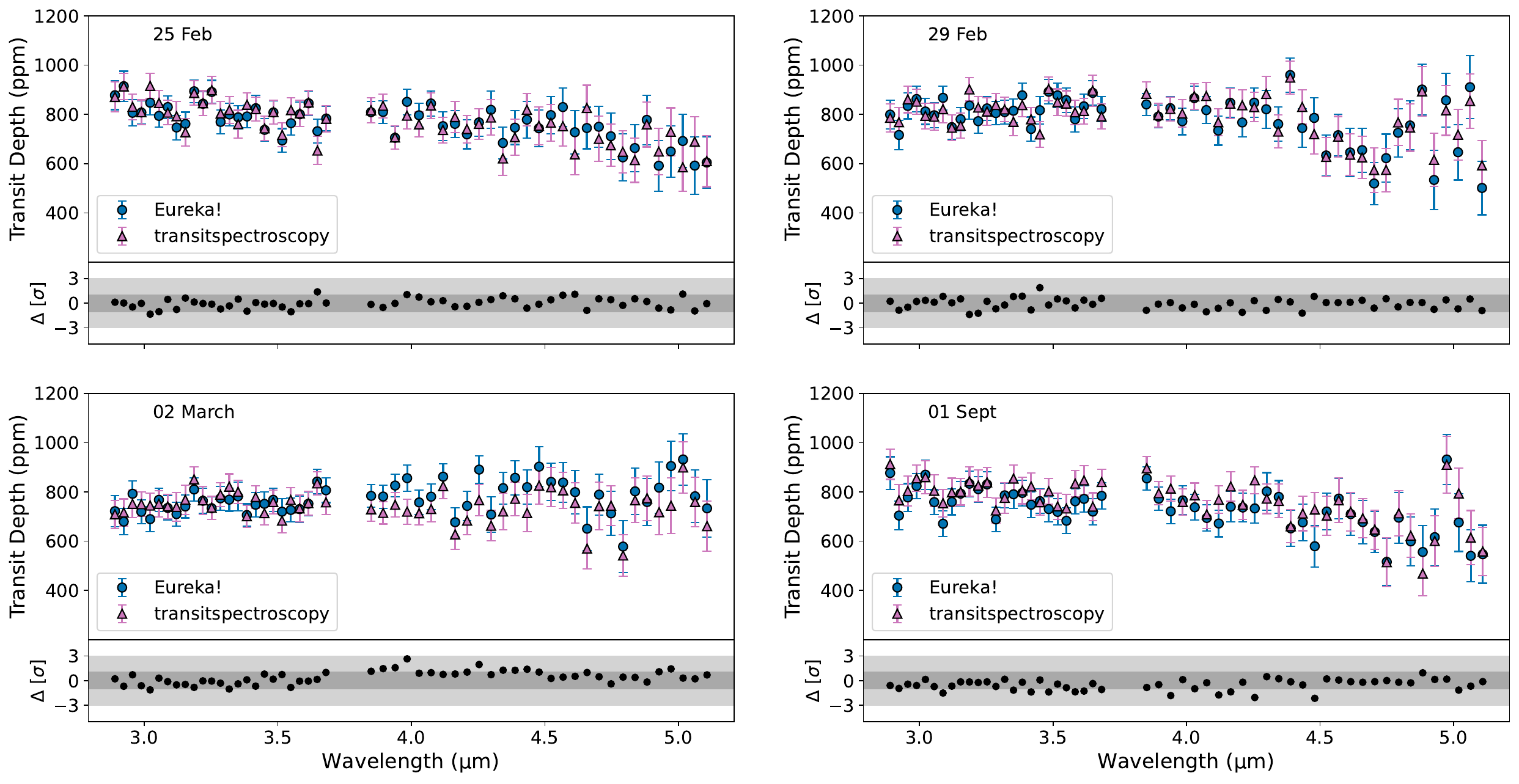}
    \caption{Comparison of the \matthewred and \tspectro reductions for our four transits. The top panels show the transit depths and their error bars. The bottom panels show the difference between the spectra, measured in $\sigma$ values relative to the \matthewred error bars.}
    \label{fig:reduction_comparison_amelie}
\end{figure*}

The individual \matthewred spectra for the four transits from our program are shown as the blue points in Figure \ref{fig:reduction_comparison_amelie}. The spectra are binned to $R\sim100$. Although there are no obvious features consistent across all four spectra, there is a notable dip around \SI{4.7}{\micron} in the 29 Feb visit. Due the size of the dip and its absence in the other visits, this is unlikely to be attributed to the planet. The significance and potential origin of this feature will be investigated and discussed in later sections. 

\begin{figure}
    \centering
    \includegraphics[width=0.5\textwidth]{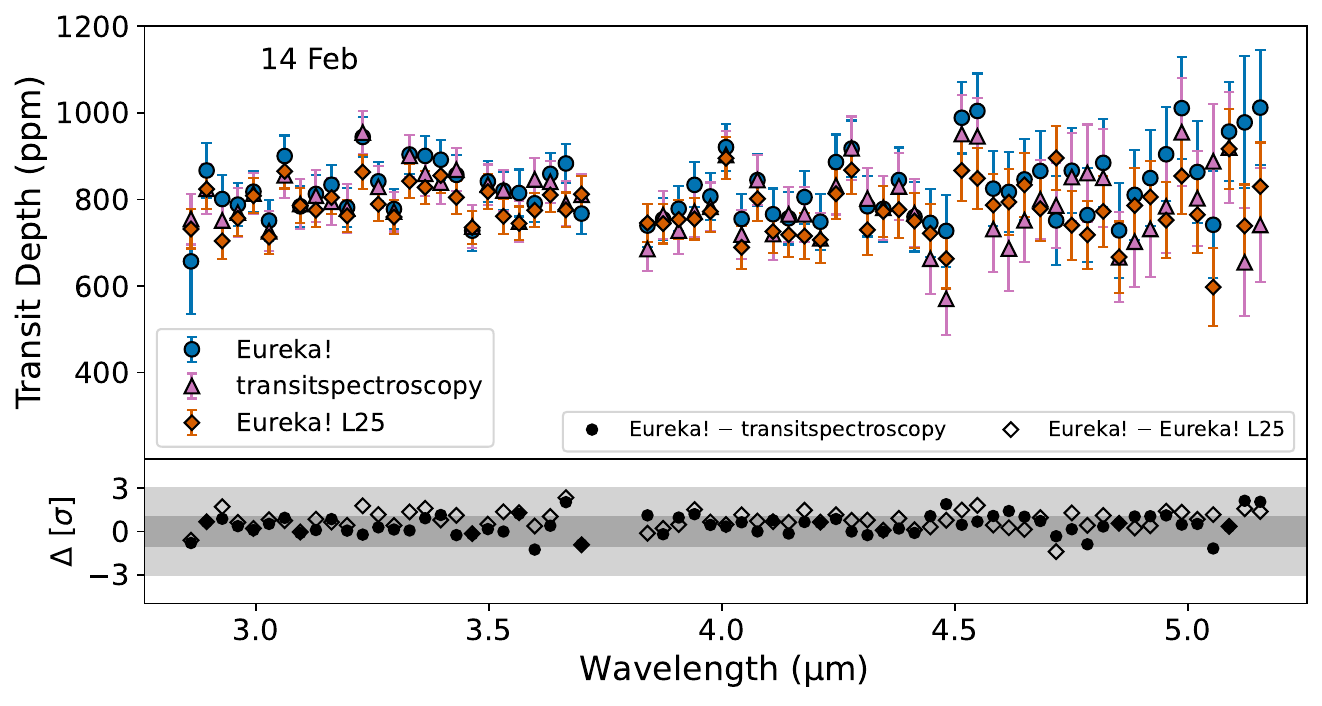}
    \caption{Comparison of the \matthewred and \tspectro reductions of the transit from the phase curve program (GO 3263) and the \texttt{Eureka!} reduction of this transit from~\protect\cite{luque25}. The top panel shows the transit depths and their error bars. The bottom panel shows the difference between the spectra, with the circles and diamonds indicating the difference between the \matthewred spectrum and the \tspectro and \protect\cite{luque25} spectra, respectively. The differences are measured in $\sigma$ values relative to the \matthewred error bars.}
    \label{fig:reduction_comparison_visit5}
\end{figure}

Figure \ref{fig:reduction_comparison_visit5} shows the \matthewred reduction of the transit from the phase curve program, compared with the reduction from \cite{luque25} (also using \matthewred) and another reduction with \tspectro (see Section \ref{sec:tspectro}). Our reductions are binned to the same data points as the spectrum from \cite{luque25} to allow for easy comparison. We find a good agreement between our \matthewred reduction and the reduction from \cite{luque25}, with all data points agreeing to within 3$\sigma$. Whilst this may not be a fully independent comparison, as both reductions are based on the \matthewred pipeline, the setups are independent, and thus the close agreement is reassuring. As with the other four transits, this visit shows no obvious spectral features that could be attributed to an atmosphere on the planet. It also shows no clear sign of the \SI{4.7}{\micron} dip seen in the 29 Feb visit. 

\begin{figure*}
    \centering
    \includegraphics[width=0.9\linewidth]{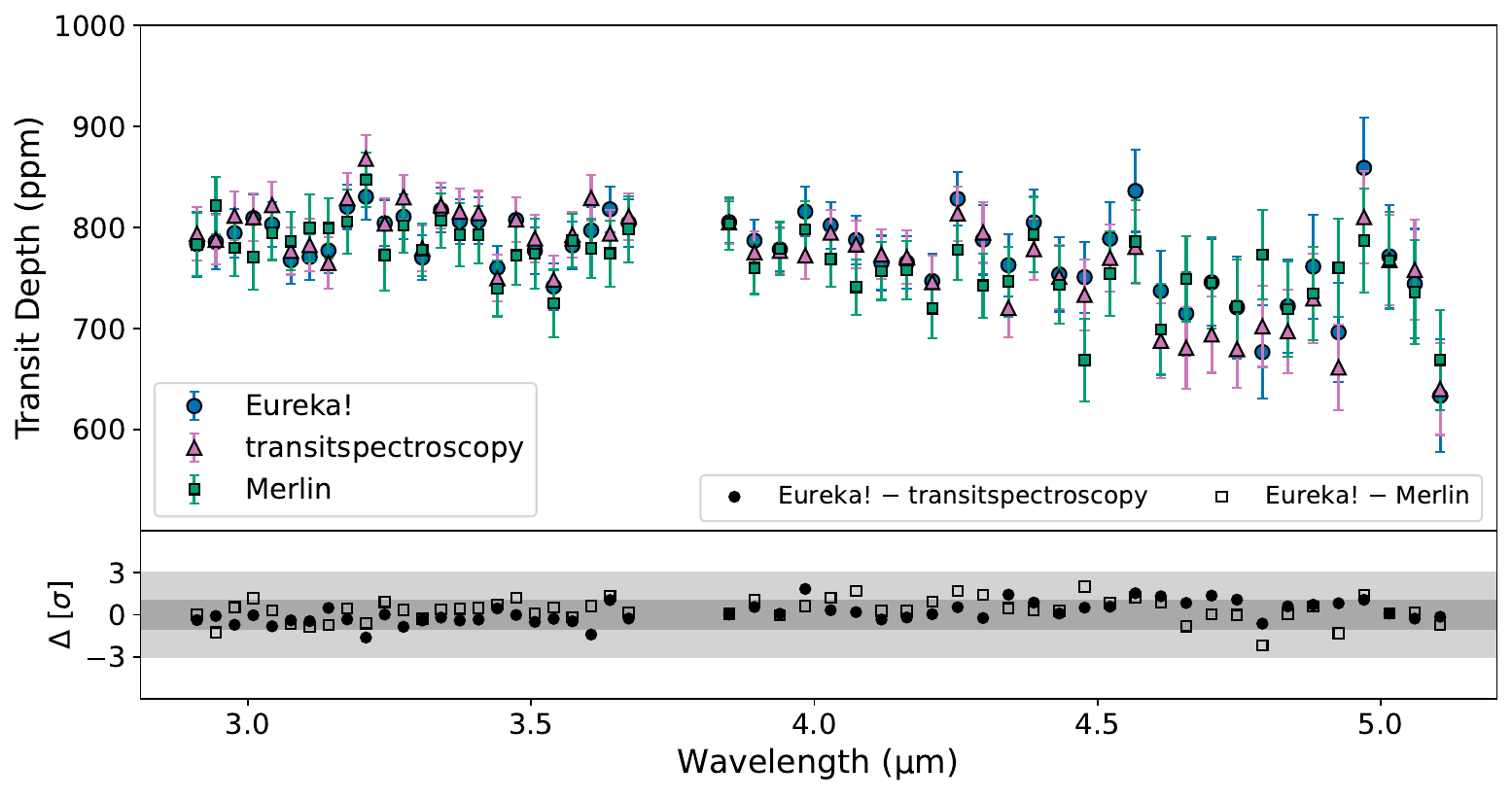}
    \caption{Comparison of the \matthewred, \tspectro and \merlinred spectra averaged over all five transits. The top panel shows the transit depths and their error bars. The bottom panel shows the difference between the spectra, measured in $\sigma$ values relative to the \tspectro and \merlinred error bars.}
    \label{fig:average_spectrum}
\end{figure*}

For the final spectrum used in our interpretation, we use the fit from our \matthewred reduction, averaged over all five transits, using an unweighted average, and binned to a resolution of $R\sim100$, after fitting the light curves, to allow comparison between the different reductions. The spectrum is shown as the blue points in Figure \ref{fig:average_spectrum}, compared with our other two reductions (see Sections \ref{sec:tspectro} and \ref{sec:merlin}). The median uncertainties for our \matthewred reduction are 23~ppm and 36~ppm for NRS1 and NRS2, respectively. All data points agree across the reductions to within 3$\sigma$ of their uncertainties. 

\subsection{Methodology 2: \tspectro}
\label{sec:tspectro}

We independently analysed the five \textit{JWST} NIRSpec transit observations of TOI-1685\,b using the \tspectro pipeline \citep{espinoza22}. Stage~1 calibration followed the standard \textit{JWST} processing steps, except we replaced the default jump detection with a custom method that calculates pixel-wise differences between consecutive groups. In Stage~2, raw integration images were converted into time-series light curves for both NRS1 and NRS2 detectors. Spectral traces were identified via cross-correlation with a Gaussian profile and smoothed using a spline. The trace spanned pixels 500--2042 (x-axis) for NRS1 and 5--2043 for NRS2. Background subtraction was performed by masking the trace in the median rate frame, computing the median of out-of-trace pixels, and subtracting this value from each integration. To mitigate 1/f noise, the median frame was scaled to each integration and subtracted. The residual column-wise 1/f noise was estimated as the median of out-of-trace pixels outside a 3-pixel inner and 8-pixel outer radius, and subtracted from each frame. Stellar flux was extracted using an aperture with a half width of 3 pixels. Pixel-level light curves were generated for each column, and white light curves were created by summing flux across all trace pixels per detector flux across all trace pixels per detector.

We fit the spectral light curves using 10-pixel bins for both detectors; light curves were not fit at the pixel level. We used \texttt{juliet} \citep{espinoza19} to model the light curves. The orbital parameters—period ($P$), mid-transit time ($t_0$), scaled semi-major axis ($a/R_\star$), and impact parameter ($b$)—were fixed to refined values from the global best-fit solution described in Section \ref{sec:eureka}. The planet-to-star radius ratio ($R_p/R_\star$) was fitted with uniform priors in the range [0, 0.2]. Quadratic limb-darkening coefficients ($u_1$, $u_2$) were sampled from uniform priors between $-3$ and $3$. Additional fitted parameters included a mean out-of-transit flux offset ($m_\mathrm{flux}$), drawn from a normal distribution with mean 0 and standard deviation 0.1, and a white noise jitter term ($\sigma_\mathrm{w}$), modelled with a log-uniform prior between 10 and 10,000~ppm. Systematic trends were modeled using an additive linear model with a single parameter, $\theta$, multiplying time (in days). A wide uniform prior between $-10$ and $10$ relative-flux~day$^{-1}$ was placed on this coefficient, as the light curves are expressed in relative flux (dimensionless) and the regressor is time in days. In addition, a Gaussian Process (GP) with a Matérn 3/2 kernel, implemented via the \texttt{celerite} package \citep{foreman-mackey17}, was used to capture correlated noise. The GP amplitude ($\mathrm{GP}\sigma$) and length scale ($\mathrm{GP}\rho$) were assigned log-uniform priors spanning $10^{-5}$ to $10^{3}$. Transit depths and their uncertainties were derived as the median and variance of the posterior distribution of $(R_p/R_\star)^2$ in each 10-pixel bin.

Figure \ref{fig:reduction_comparison_amelie} shows the \tspectro spectra for the four transits of our program, compared with the reductions from \matthewred. There is a close agreement between the two reductions in all visits, with all data points within 3$\sigma$ of each other, and most within 1$\sigma$. Figure \ref{fig:reduction_comparison_visit5} shows the comparison for the transit from the phase curve observation. Again, the agreement is very good. Finally, the average of all five visits is shown in Figure \ref{fig:average_spectrum}, showing consistency between the different reductions. The median uncertainties for our \tspectro average spectrum are 24~ppm and 33~ppm for NRS1 and NRS2, respectively.

\subsection{Methodology 3: \merlinred}
\label{sec:merlin}

For this data reduction, we followed the recommended JWST pipeline steps using CRDS version 1.1.16 reference files \citep{bushouse25} up to the 1/f correction stage. Specifically, we applied the steps \textit{gain\_scale}, \textit{dq\_init}, \textit{saturation}, \textit{superbias}, \textit{refpix}, \textit{linearity}, and \textit{dark\_current}, with thresholds set by the corresponding CRDS reference files.

We performed the 1/f noise correction by subtracting the column-wise median outside the trace mask. To do this, we first fit a quadratic trend to the spectral trace and then defined an exclusion radius around it, fixed to the same value for each column. To determine this radius for each visit, we selected the radius that minimized the root mean square (RMS) of residuals in a crude out-of-transit (OOT) white light curve fit. Averaging these optimal radii across all visits, we adopted a final value of 9 pixels for both NRS1 and NRS2.

To detect jumps, we applied a $4\sigma$ rejection threshold. During \textit{ramp\_fit}, we excluded the final group because, in some integrations, the linearity correction slightly overcompensates the last group’s decrease in counts. We applied \texttt{weighted\_fit} to all integrations. After ramp fitting, we performed a secondary background subtraction similar to the group-wise 1/f correction. This step further reduced the OOT RMS noise. Again, we found that using a 9-pixel exclusion radius across all wavelengths and visits yielded the lowest RMS.

For aperture extraction, we modeled the background relative to the same quadratic fit to the spectral trace that was used in the 1/f correction step. We optimized the aperture radii by minimizing the OOT RMS of the final white light curves, selecting 6 pixels for NRS1 and 5 pixels for NRS2.

To compare our results with the other methodologies, we used 10-pixel spectral bins for light curve fitting. After normalising the spectral light curves, we propagated the weighted errors by summing in quadrature across spectral columns and dividing by the square root of the bin size.

For each spectral bin, we fit a transit model using \texttt{juliet} \citep{espinoza19}. We held the parameters $P$, $T_0$, $a/R_\star$, $b$, fixed to the values derived from the global white light curve fit. For the limb-darkening parameters $u_1$ and $u_2$, we used Gaussian priors centered on coefficients computed from \textit{ExoTiC-LD} tables \citep{grant24} and the MPS-ATLAS-1 database \citep{kostogryz22}, with standard deviation of 0.1 and truncated to the range $(-3,3)$.

We fit the planet-to-star radius ratio ($R_p/R_\star$) independently in each bin and modeled correlated noise using a GP with a squared-exponential kernel as implemented in \texttt{juliet} (via \texttt{celerite} \citep{foreman-mackey17}). The GP used hyperparameters $\sigma_\mathrm{GP}$ and $\rho_\mathrm{GP}$, representing the amplitude and timescale of the kernel, respectively. We placed log-uniform priors on these hyperparameters: $\sigma_\mathrm{GP} \sim \mathrm{LogUniform} (10^{-5}, 10^4)$ and $\rho_\mathrm{GP} \sim \mathrm{LogUniform} (10^{-4}, 3)$.

\begin{figure*}
    \centering
    \includegraphics[width=\textwidth]{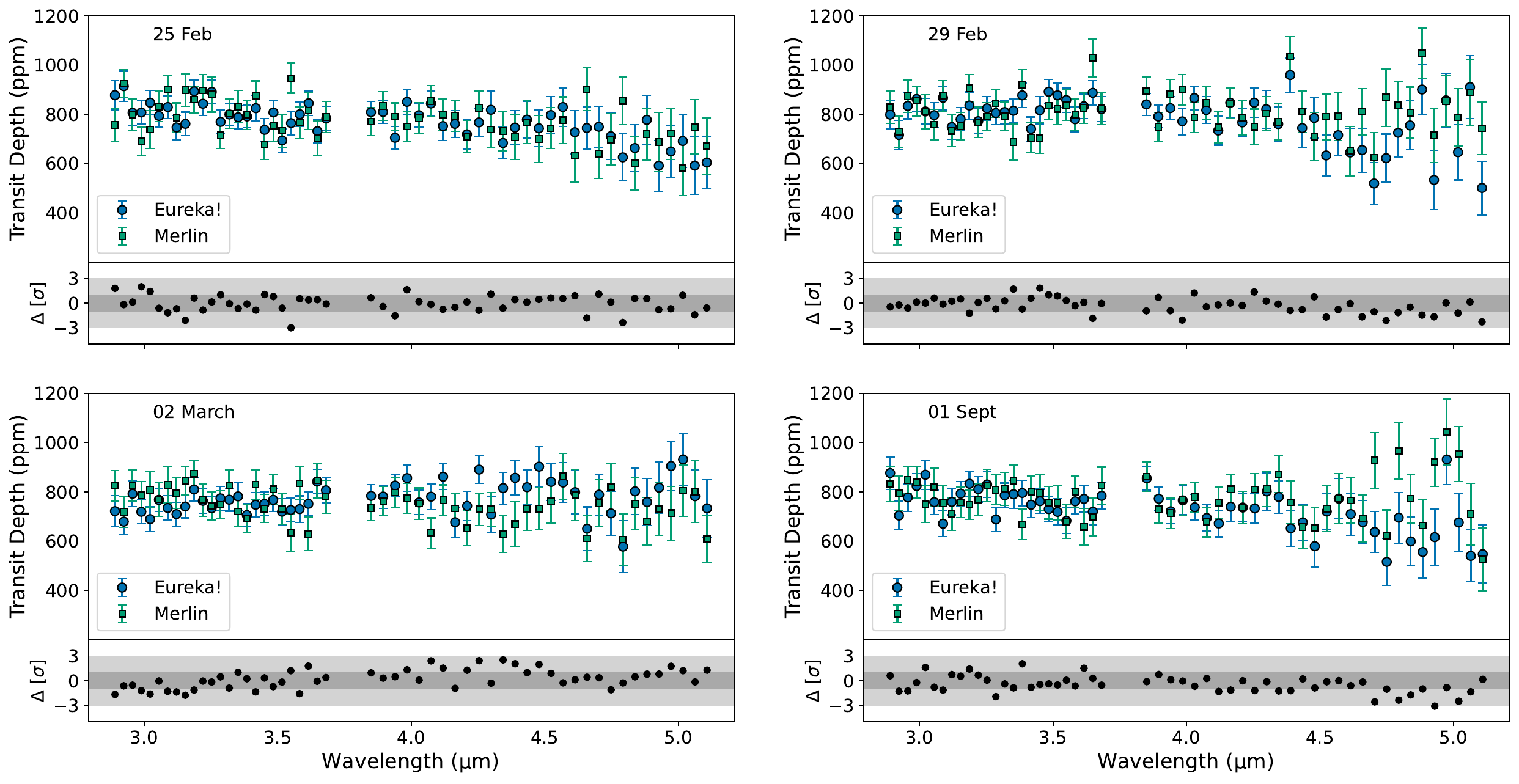}
    \caption{Comparison of the \matthewred and \merlinred reductions for our four transits. The top panels show the transit depths and their error bars. The bottom panels show the difference between the spectra, measured in $\sigma$ values relative to the \merlinred error bars.}
    \label{fig:reduction_comparison_merlin}
\end{figure*}

The \merlinred spectra for each of the four transits in our program are shown in Figure \ref{fig:reduction_comparison_merlin}, compared to the \matthewred spectra\footnote{Figures \ref{fig:reduction_comparison_amelie} and \ref{fig:reduction_comparison_merlin} display only the pairwise comparisons of each reduction with the \matthewred reduction, for clarity.}. There is a good agreement between the different reductions, with all points agreeing within $\sim3\sigma$ of their uncertainties. Perhaps the most notable difference is the lack of a distinct dip at \SI{4.7}{\micron} in the \merlinred reduction of the 29 Feb visit. The average of all five transits of the \merlinred reduction is shown in Figure \ref{fig:average_spectrum}. Despite some minor disagreement in the individual visits, the average spectrum shows a closer agreement between the different reductions. The median uncertainties for the \merlinred reduction are 29~ppm and 41~ppm for NRS1 and NRS2, respectively.

\section{Analysis}
\label{sec:analysis}

With a good agreement between our data reduction methods, we can now perform our analyses, to assess whether any potential features in the data may have a physical origin or can be meaningfully interpreted. For this analysis, we use the \matthewred reduction, averaged over all five visits and binned to a resolution of $R\sim100$. Recent studies of JWST transmission spectra of terrestrial planets have established a best-practice for analysing potentially flat spectra \citep{alderson24,alderson25,alam25,scarsdale24}. This involves first performing a comparison between flat line models with those including a simple Gaussian feature added on top, to establish the significance of any possible features without biasing towards particular molecules. Secondly, in the event of a featureless spectrum, physical atmospheric models can be used to determine the parameter space that can be ruled out by the data, namely considering the limits on metallicity and cloud deck. In the subsequent sections we follow this methodology to analyse our transits of TOI-1685~b. 

\subsection{Gaussian Tests}
\label{sec:gaussian_tests}

Following \citet{alderson25}, we test a set of different synthetic models on each of our individual transmission spectra: 1) a flat line, 2) two flat lines with an offset between NRS1 and NRS2, 3) a two-parameter sloped line, 4) a flat line with a Gaussian feature, and 5) a flat line with a negative Gaussian feature. The aim of these tests is to establish whether there is statistically significant structure in the data before committing to interpreting it with physically motivated models. Though the fifth model is not a feature we would expect from a rocky planet's atmosphere in transmission, by-eye we appear to have a strong dip in the spectrum around \SI{4.7}{\um} in the 29 Feb transit, and we would like to determine its significance. We can also see hints of a slope in the spectra for the 25 Feb and 01 Sept transits, and the third model will test the significance of these. 

By comparing the Bayesian evidences and reduced-$\chi^2$ of the five models' fit to the data, we can determine if any statistically significant Gaussian features are present in the spectra. For two models, the Bayes factor (i.e. the ratio of the models' Bayesian evidences) can be used to quantify the evidence in favour of one model over the other. For reference, the Jeffreys' scale gives the log-Bayes factor\footnote{$\ln{B} = \Delta\ln{Z}$} $\ln{B}<1.0$ as inconclusive evidence, $1.0<\ln{B}<2.5$ as weak evidence, $2.5<\ln{B}<5.0$ as moderate evidence, and $\ln{B}>5.0$ as strong evidence in favour of one model over another \citep[e.g.,][]{trotta08}. We perform these fits to each of the five transmission spectra individually, and to the averaged spectrum, for each of the three data reductions. The results are shown in Table \ref{tab:gaussian_test}, with the lowest-complexity model favoured by the data (i.e. $\ln{B}<1$) shown in bold. 

\bgroup
\def\arraystretch{1.3}%
\begin{table*}
    \centering
    \begin{tabular}{|c|cc|cc|cc|cc|cc|cc|}
        \hline
         & \multicolumn{2}{c|}{14 Feb} & \multicolumn{2}{c|}{25 Feb} & \multicolumn{2}{c|}{29 Feb} & \multicolumn{2}{c|}{02 March} & \multicolumn{2}{c|}{01 Sept} & \multicolumn{2}{c|}{Average} \\
        Model (free parameters) & $\Delta\ln{Z}$ & $\chi^2/\nu$ & $\Delta\ln{Z}$ & $\chi^2/\nu$ & $\Delta\ln{Z}$ & $\chi^2/\nu$ & $\Delta\ln{Z}$ & $\chi^2/\nu$ & $\Delta\ln{Z}$ & $\chi^2/\nu$ & $\Delta\ln{Z}$ & $\chi^2/\nu$ \\
        \hline
        Flat Line (1) & \textbf{0.00} & \textbf{1.17} & 4.65 & 1.04 & 4.14 & 1.21 & \textbf{0.00} & \textbf{0.87} & 5.25 & 1.17 & 2.36 & 1.27 \\
        Offset (2) & 2.47 & 1.17 & 4.76 & 0.94 & 5.31 & 1.15 & -0.82 & 0.73 & 5.03 & 1.06 & 3.40 & 1.20 \\
        Slope (2) & 2.40 & 1.19 & \textbf{0.00} & \textbf{0.75} & 3.52 & 1.10 & -1.55 & 0.73 & \textbf{0.00} & \textbf{0.88} & \textbf{0.00} & \textbf{1.09} \\
        Flat Line + Gaussian (4) & 0.27 & 1.08 & 3.52 & 0.86 & 3.48 & 1.00 & -0.42 & 0.74 & 4.62 & 1.01 & 2.27 & 1.14 \\
        Flat Line + Negative Gaussian (4) & 0.20 & 1.12 & 2.08 & 0.80 & \textbf{0.00} & \textbf{0.86} & -0.47 & 0.73 & 2.57 & 0.91 & -0.53 & 1.02 \\
        \hline
    \end{tabular}
    \caption{$\Delta\ln{Z}$ and $\chi^2/\nu$ values for each model, listed in the first column, fit to each of the \matthewred transits, and the average spectrum, at a resolution of $R\sim100$. $\Delta\ln{Z}$ is compared to the model with the fewest free parameters favoured by the data (i.e. with $\Delta\ln{Z}<1.0$), which is highlighted in bold. In other words, the model with the lowest complexity is favoured unless a more complex model has $\Delta\ln{Z}>1.0$. $\nu$ denotes the degrees of freedom, equal to the number of data points minus the number of free parameters in the model, given in brackets in the first column.}
    \label{tab:gaussian_test}
\end{table*}
\egroup

For all transits, including the average, there is no evidence for the flat line + Gaussian model, which would be indicative of potential spectral features and an atmospheric detection. However, as seen by-eye there is some evidence for the flat line + negative Gaussian model in the 29 Feb transit (log-Bayes factor $\ln{B}=4.14$ and $\ln{B}=3.52$ compared to the flat line and slope models, respectively, corresponding to moderate evidence). A negative Gaussian feature could be caused by instrument systematics, or possibly stellar activity, which we will discuss further in Section \ref{sec:discussion}. Since the fits to the average spectrum and the other transits do not find evidence for this feature, we proceed without further action. 

For two of the transits (25 Feb and 01 Sept) we find some evidence for the slope model. This could have a number of physical explanations, from an atmosphere with clouds/hazes, to stellar activity, to instrument systematics. However, the log-Bayes factor comparison with the flat line or offset models is at most about 5, corresponding to the limit for strong evidence, and in the average spectrum case is only 2.36, constituting weak evidence. Though the results are only shown for the \matthewred reduction, the analysis of the \tspectro and \merlinred reductions have qualitatively comparable results. The main difference is that the \tspectro and \merlinred reductions show comparatively more and less evidence, respectively, for the negative feature at \SI{4.7}{\um} than the \matthewred reduction. 

These results show minimal evidence for anything beyond a flat spectrum with no signs of an atmosphere. Despite the lack of spectral features, this data can still be useful for ruling out different regions of parameters space for atmospheric models, and we will explore this in detail in the following sections. 

\begin{figure}
    \centering
    \includegraphics[width=0.5\textwidth]{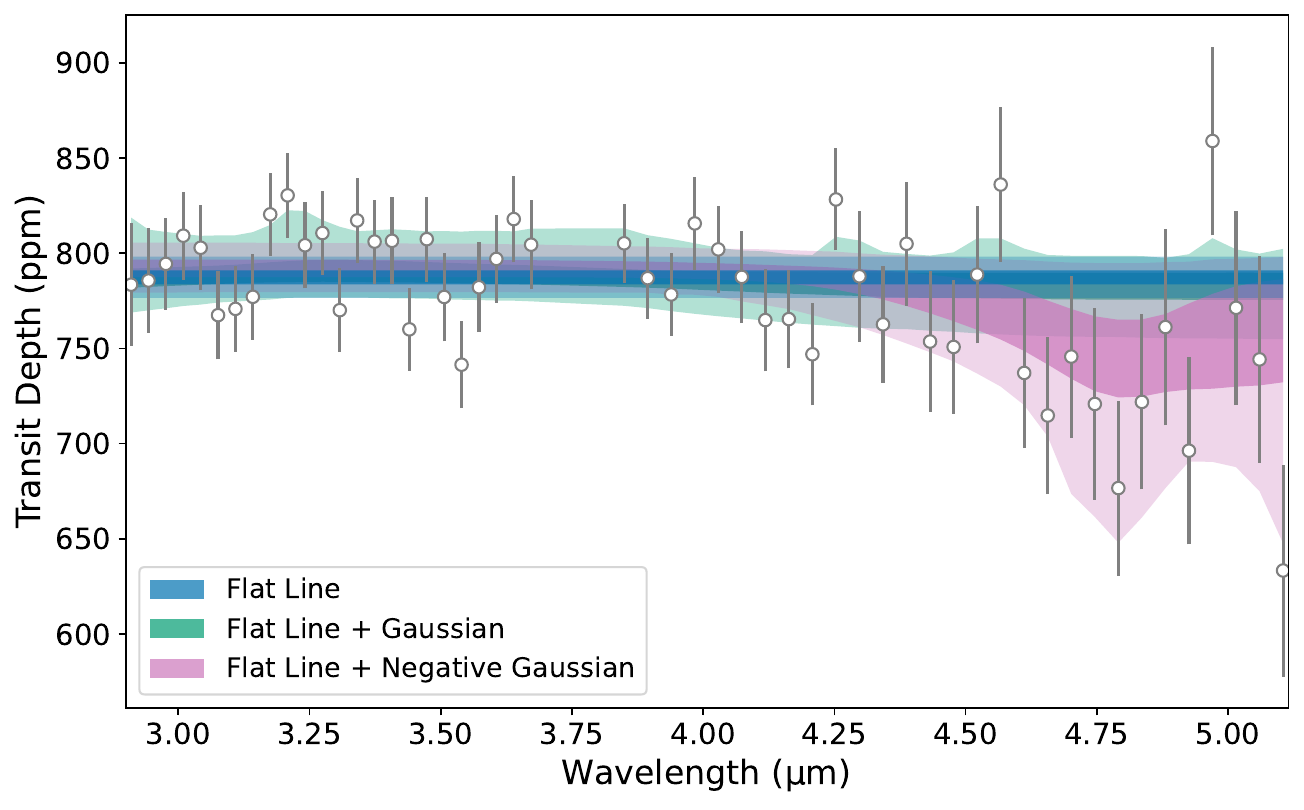}
    \caption{Fits to the \matthewred spectrum, averaged over all five transits, using the analytical models. The dark and pale shaded regions show the 1- and 3-$\sigma$ fits, respectively. The log-Bayes factor comparison between the flat line (blue) and the flat line plus a Gaussian (green) is 0.09, equivalent to inconclusive evidence. The log-Bayes factor comparison between the flat line (blue) and the flat line plus a negative Gaussian (pink) is 2.89, equivalent to moderate evidence.}
    \label{fig:gaussian_test}
\end{figure}

\subsection{Atmospheric Modelling}
\label{sec:modelling}

\subsubsection{Chemical Equilibrium}

Again, following \citet{alderson25}, we also compare the spectrum to a range of physical atmospheric models. The goal is to determine the region of parameter space that can be confidently ruled out by the data. We start by considering chemical equilibrium models. Although we may not expect a rocky planet's atmosphere to be in chemical equilibrium, this is nevertheless a useful baseline test to determine possible atmospheric mean molecular weights and thicknesses allowed by the data. 

For these models, we use the forward model from the open-source retrieval code \texttt{BeAR}\footnote{Formerly known as \texttt{Helios-r2}. The code can be found here: \url{https://github.com/newstrangeworlds/bear}} \citep{kitzmann20}. We set the temperature profile of the atmosphere to be isothermal, and equal to the planet's equilibrium temperature of 1062 K. The atmosphere is divided into 200 levels, equal in log-space, and ranging from 10 bar to $10^{-8}$ bar. We include absorption from \ch{H2} Rayleigh scattering \citep{cox00}, as well as collision-induced absorption from \ch{H2}-\ch{H2} \citep{abel11} and \ch{H2}-\ch{He} pairs \citep{abel12}, computed from the \texttt{HITRAN} database \citep{karman19}. For molecular absorption, we include the following species and their associated line-lists from \texttt{ExoMol}: \ch{H2O} \citep{polyansky18}, \ch{CO} \citep{li15,somogyi21}, \ch{CO2} \citep{yurchenko20}, \ch{CH4} \citep{yurchenko13,yurchenko14}. The corresponding opacities have been calculated using the HELIOS-K code \citep{grimm15, grimm21} and are available through the DACE platform\footnote{\url{https://dace.unige.ch/}}. 

To compute the mixing ratios of each species, \texttt{BeAR} uses the open-source chemical equilibrium code \texttt{FastChem}\footnote{\url{https://github.com/newstrangeworlds/fastchem}} \citep{stock18,stock22,kitzmann24}. For our models, we vary the atmospheric metallicity, but fix the C/O ratio to the solar value (0.55). We adopt the solar elemental abundances from \cite{asplund09}. We also include a grey cloud at varying pressure levels. This acts as an agnostic opaque pressure level, which could correspond to either a cloud deck or the planet's surface. 

\begin{figure*}
    \centering
    \includegraphics[width=0.9\textwidth]{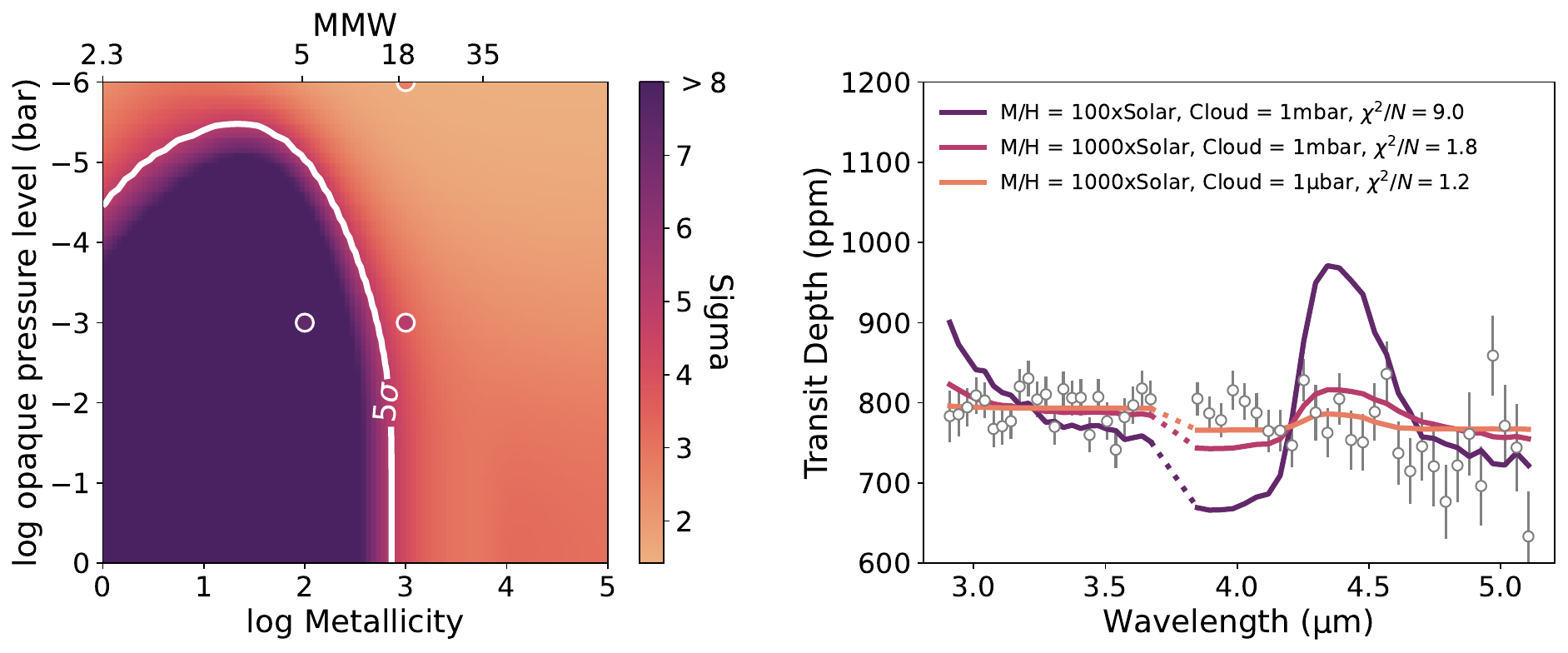}
    \caption{Comparison between the \matthewred average spectrum and a variety of atmospheric models assuming chemical equilibrium, spanning a range of metallicities and opaque pressure levels. The left panel shows the $\sigma$-confidence for ruling out different regions of parameter space. The right panel shows three example equilibrium models along with the data. The corresponding parameter values for these models are shown as the white circles on the left panel.}
    \label{fig:equilibrium_fit}
\end{figure*}

To explore the parameter space we vary the metallicity from solar to $10^5\times$ solar, and vary the opaque pressure level from 1 bar to $10^{-6}$ bar, both with 100 equal steps in log-space. This results in 10,000 models, each of which is then shifted to fit the data, allowing for an offset between NRS1 and NRS2. Finally, we perform a $\chi^2$ test to determine the confidence with which we can rule out the model. Figure \ref{fig:equilibrium_fit} shows the results for this test. Almost all models with metallicities below  $\sim1000\times$ solar can be confidently ruled out by the data. The exception occurs at very low opaque pressure levels, however these values may not be physically plausible. If the opaque pressure level is interpreted as a cloud deck, this requires very high-altitude aerosols in the atmosphere. Clouds forming through condensation of trace gases would need to be lofted to low pressures, as they form deeper in the atmosphere \citep{ackerman01,gao18,powell18}. This requires extremely strong vertical mixing and is unlikely to be physical \citep[e.g.,][]{gao18a}. Photochemical hazes can form at lower pressures, but this leads to small particles that would not explain the flat spectra at these longer wavelengths \citep[e.g.,][]{morley15,gao20,gao21}. If instead the opaque pressure level refers to the surface pressure, this would imply an extremely thin atmosphere, which would be highly susceptible to atmospheric escape (see Section \ref{sec:escape}).

Higher metallicities result in sufficiently flat spectra that are not ruled out by the data, due to their high mean molecular weight. At equilibrium, and with a solar C/O ratio, this corresponds to $\mu\gtrsim$18, with lower mean molecular weights ruled out to a significance of $\sim5\sigma$, for opaque pressure levels of at least 1~mbar. The offset between NRS1 and NRS2 varies between 0 and -100ppm, with models containing larger spectral features requiring more extreme offsets, as seen in the right panel of Figure \ref{fig:equilibrium_fit}.

\subsubsection{\ch{H2}-Molecule Mixtures}
\label{sec:h2_mixtures}

Next we consider single molecule mixtures with \ch{H2} backgrounds. For this, we use the same set up of the \texttt{BeAR} forward model as before, but without the chemical equilibrium assumption. Instead, we take a single molecule and vary its volume mixing ratio from $10^{-6}$ to 1.0, at 100 equal steps in log space. We then fill the remainder of the atmosphere with \ch{H2}/\ch{He} at a solar ratio. We perform the same opaque pressure level variations as before. 

\begin{figure*}
    \centering
    \includegraphics[width=0.9\textwidth]{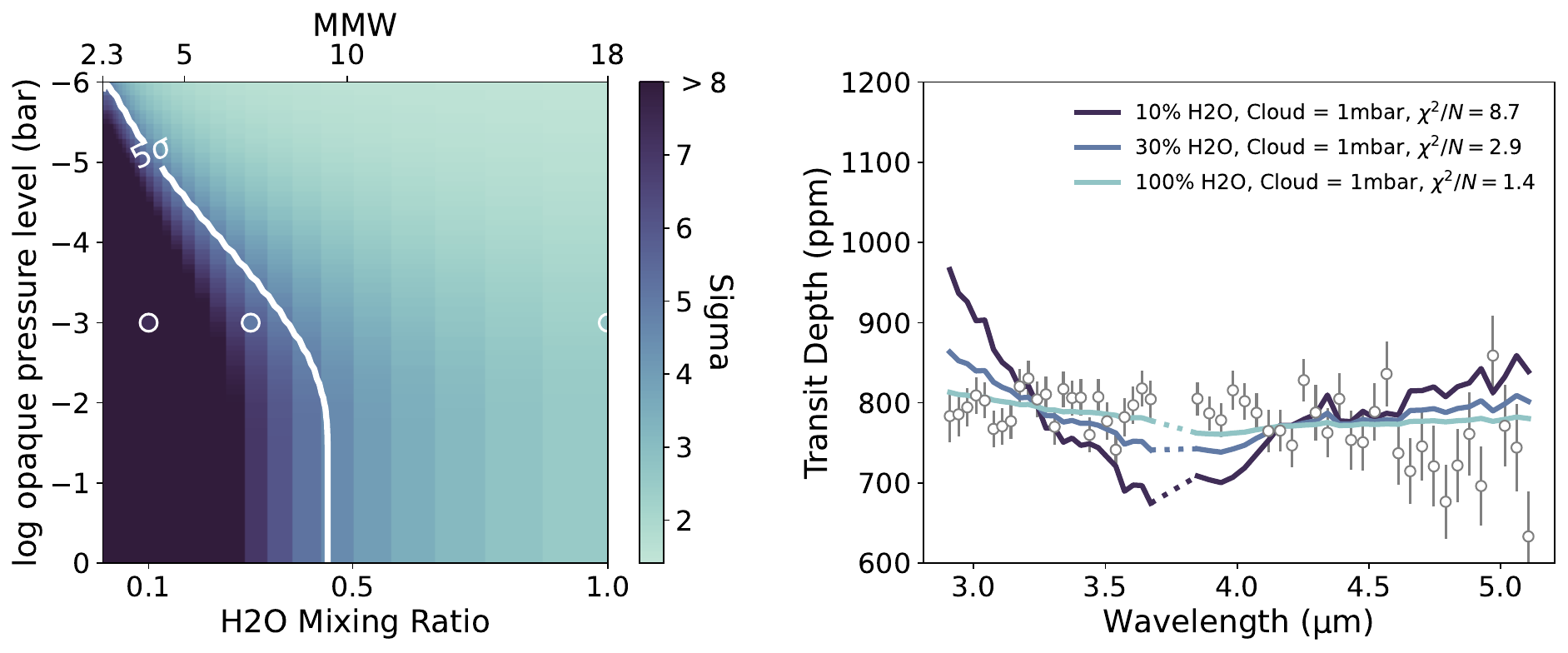}\\
    \vspace{0.3cm}
    \includegraphics[width=0.9\textwidth]{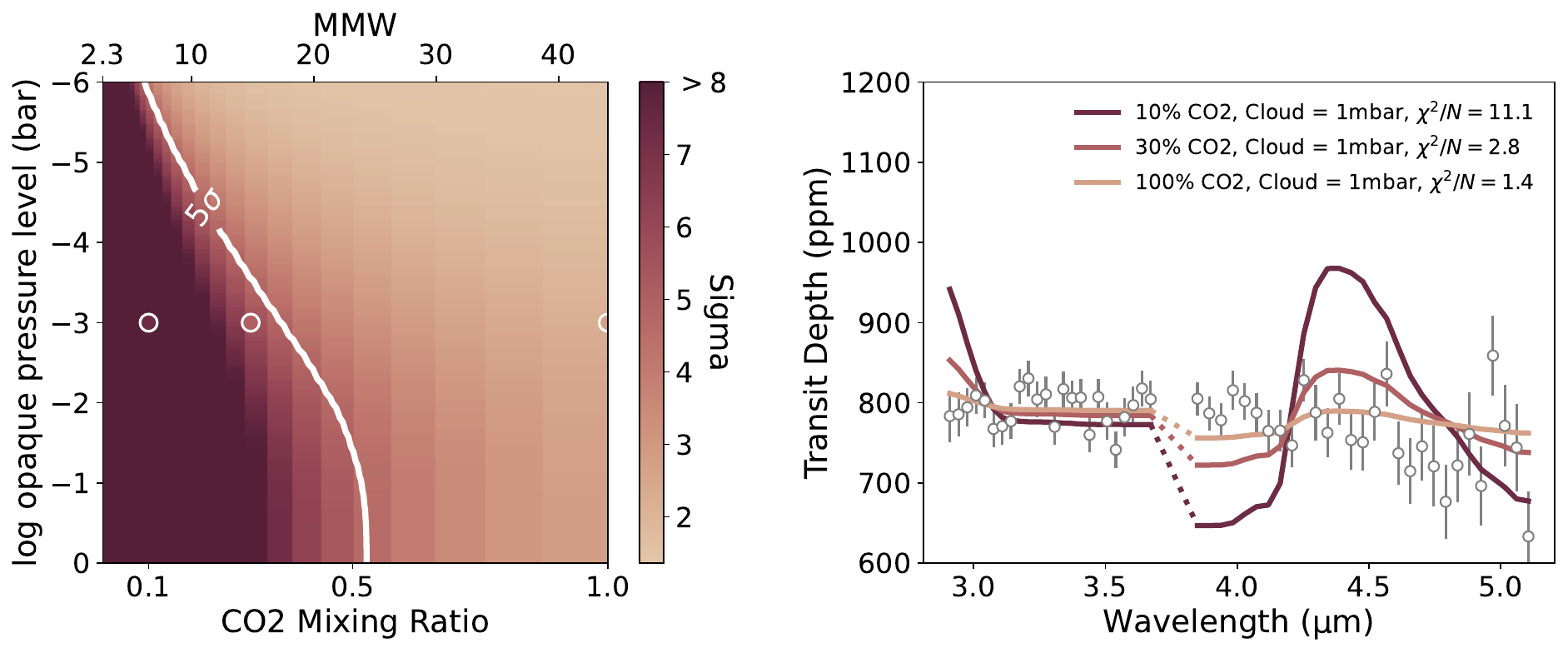}
    \caption{Comparison between the \matthewred spectrum and a variety of atmospheric models assuming an \ch{H2}-molecule mixture, spanning a range of molecule mixing ratios and opaque pressure levels. The top and bottom panels show the results for \ch{H2O} and \ch{CO2}, respectively. In each case, the left panel shows the $\sigma$-confidence for ruling out different regions of parameter space, and the right panel shows three example models along with the data. The corresponding parameter values for these models are shown as the white circles on the left panels.}
    \label{fig:H2O_CO2_fit}
\end{figure*}

Figure \ref{fig:H2O_CO2_fit} shows the results for \ch{H2O} and \ch{CO2}. For \ch{H2O}, we are able to rule out mixing ratios below $\sim50$\% for opaque pressure levels $\gtrsim1$mbar, corresponding to $\mu\gtrsim$10. Again, very low opaque pressures would enable much lower molecular weights, but may not be physically plausible (see Section \ref{sec:escape}). The NRS offset varies between -20 and 190ppm, with models containing larger water features requiring a larger offset.

For \ch{CO2}, we can rule out mixing ratios below $\sim50$\% for opaque pressure levels $\gtrsim10$mbar. Due to its heavier molecular weight, this corresponds to $\mu\gtrsim$20. In the \ch{CO2} case, a mean molecular weight below 10 would require an opaque pressure level $\lesssim$\SI{10}{\micro\bar}, which is unlikely to be sustained. The NRS offset varies between -20 and -950ppm, again with models containing larger \ch{CO2} features requiring a larger offset. Note that these offset values are not necessarily realistic, as we did not impose physically-motivated prior limits.

\begin{figure*}
    \centering
    \includegraphics[width=0.9\textwidth]{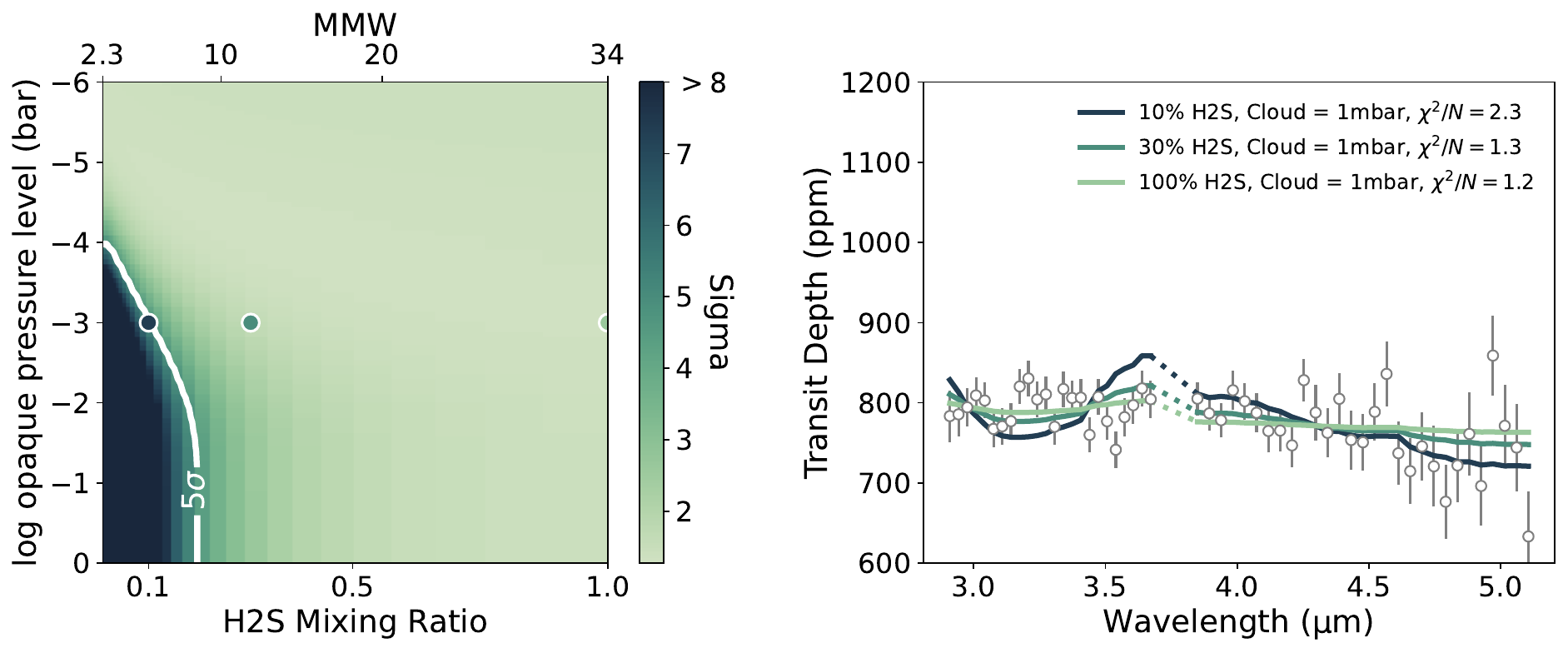}\\
    \vspace{0.3cm}
    \includegraphics[width=0.9\textwidth]{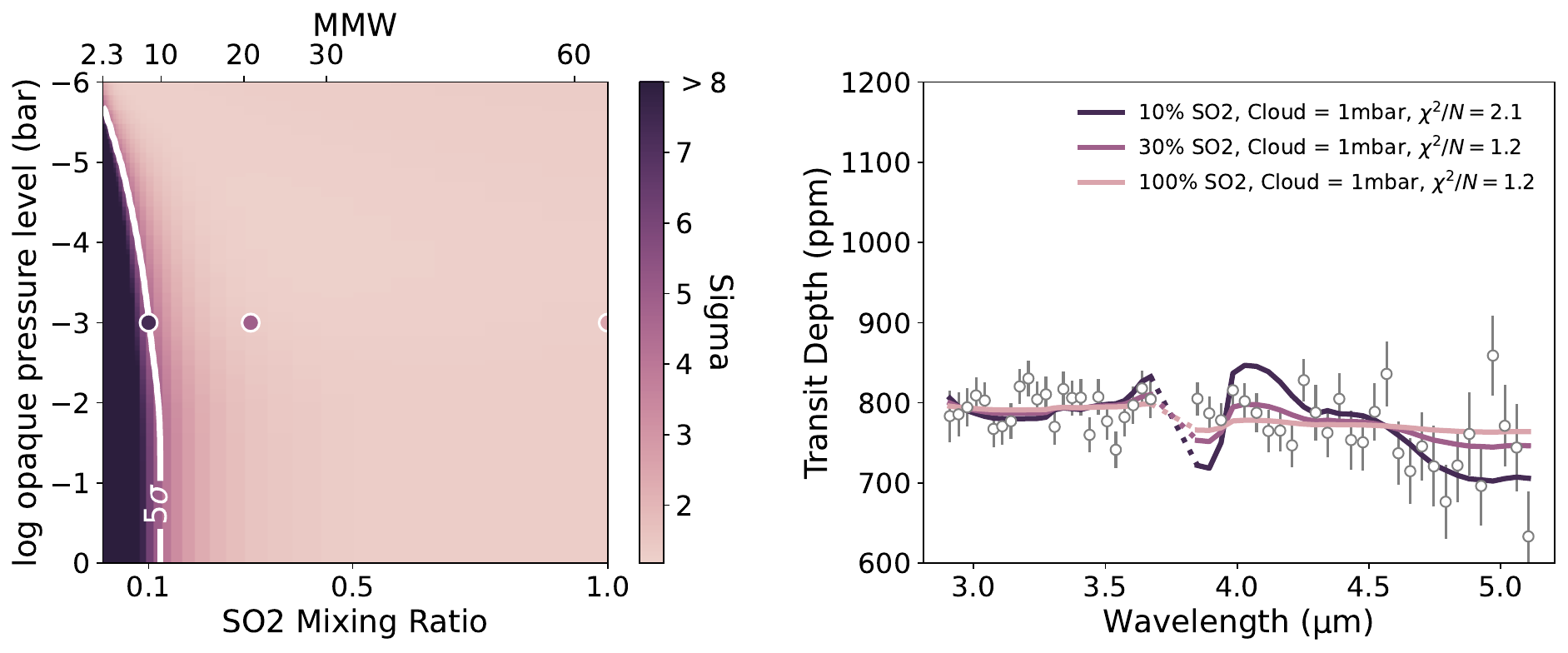}
    \caption{Comparison between the \matthewred spectrum and a variety of atmospheric models assuming an \ch{H2}-molecule mixture, spanning a range of molecule mixing ratios and opaque pressure levels. The top and bottom panels show the results for \ch{H2S} and \ch{SO2}, respectively. In each case, the left panel shows the $\sigma$-confidence for ruling out different regions of parameter space, and the right panel shows three example models along with the data. The corresponding parameter values for these models are shown as the white circles on the left panels.}
    \label{fig:H2S_SO2_fit}
\end{figure*}

We also consider the possibility of sulfur species, which may have been detected in L 98-59 d \citep{banerjee24,gressier24}, and could be a result of volcanism on the planet. Figure \ref{fig:H2S_SO2_fit} shows the results for \ch{H2} backgrounds with \ch{H2S} and \ch{SO2}. As these molecules also have high molecular weights, the limits on their mixing ratios are more similar to \ch{CO2}. For \ch{H2S}, we can rule out mixing ratios below $\sim20\%$, and for \ch{SO2} the limit is $\sim10\%$. The NRS offsets vary between -20 and -140ppm for \ch{H2S}, and between -20 and -690ppm for \ch{SO2}, again with models containing larger spectral features requiring larger offsets.

These limits generally correspond to a lower limit on the mean molecular weight of the atmosphere of $\sim10$ at $\sim5\sigma$, for a minimum opaque pressure level of $\sim1$-$10$~mbar. Higher mean molecular weight atmospheres cannot be ruled out from this analysis. 

\subsection{Atmospheric Retrievals}
\label{sec:retrievals}

Due to the lack of obvious spectral features, a full atmospheric retrieval exploration is not warranted by the data. However, \cite{luque25} used a small set of retrievals to test the ability of their transmission spectrum to rule out certain atmospheric scenarios, in a similar vein to Section \ref{sec:modelling}. In order to directly compare the gain in constraining power from obtaining four additional transits of the planet, we perform the same set of retrievals on our average spectrum. As in Section \ref{sec:modelling} we use \texttt{BeAR}, this time making use of its retrieval capabilities. \texttt{BeAR} uses the \texttt{MultiNest} library \citep{feroz08} to perform Bayesian nested-sampling \citep{skilling04}. In our retrievals, we use 1000 live points, and an evidence tolerance of 0.5.

As in \cite{luque25}, we start with a hydrogen-dominated atmosphere, with 1\% \ch{H2O}, and test this with and without clouds. We also consider a cloud-free hybrid atmosphere with 90\% \ch{H2} and 10\% \ch{H2O}. We then have five potential secondary atmospheres consisting of pure \ch{CO2}, \ch{SO2}, \ch{H2O}, or \ch{CH4}, and a mixed case with \ch{CO}+\ch{CO2}+\ch{SO2}, all including clouds. For each retrieval, the only free parameters are the planet radius (i.e. the continuum level), the detector offset, and, if included, the cloud-top pressure. The abundances are fixed, and the temperature profile is isothermal and set to the equilibrium value. The exception is the mixed secondary atmosphere case, where the abundances of \ch{CO} and \ch{CO2} are also retrieved, using centred-log-ratio priors \citep{benneke12}, and the abundance of \ch{SO2} then fills the remaining atmosphere. 

The results of these retrieval tests are shown in Table \ref{tab:retrievals}. The evidence of each model is compared relative to the flat line (including a detector offset). In \cite{luque25}, their retrievals on a single transit were only able to confidently rule out the clear \ch{H2}-dominated and hybrid atmospheres. However, they found that the cloudy \ch{H2}-dominated atmosphere was only possible in the case of an extremely high cloud deck and a very small planet radius. In comparison, we are able to very confidently rule out both \ch{H2}-dominated scenarios and the hybrid case. We find that both the pure \ch{H2O} and \ch{CO2} scenarios are only moderately or weakly ruled out, but require a high cloud deck, above 1~mbar and 0.1~mbar, respectively (note that this could also be the atmospheric thickness above a surface). In \citet{luque25}, a pure \ch{CH4} atmosphere was found to have a marginally higher Bayesian evidence than the flat line. However, in our results we find it to be weakly ruled out by the data. Furthermore, as noted in \cite{luque25}, a pure \ch{CH4} atmosphere is a physically unlikely scenario, due to photochemical effects \citep[e.g.,][]{moses11}. Finally, both the pure \ch{SO2} and mixed secondary atmospheres are possible, with or without clouds, consistent with \citet{luque25}.

From this set of atmospheric retrievals, we find no evidence for the presence of an atmosphere on TOI-1685~b, though thin secondary atmospheres, or those with a high cloud deck, cannot be completely ruled out. This appears to be in contrast with Section \ref{sec:h2_mixtures}, where the \ch{H2}-mixtures containing high percentages of other molecules are not ruled out for any opaque pressure level. However, the goal of the retrievals is to determine the statistical preference for the flat line compared to other physical models. Overall, the combination of analyses indicates that, whilst the secondary atmospheres can obtain a reasonable fit to the data, the model complexity is not warranted, and, invoking Occam's razor, we conclude that the spectrum is best explained by a flat line.

\bgroup
\def\arraystretch{1.5}%
\begin{table*}
    \centering
    \begin{tabular}{|c|c|c|c|}
        \hline
         Model & $\ln{B}$ & Consistent? & Remarks \\
         \hline
         Flat line & -- & Y & No chemistry. \\
         Clear primary atm. (99\% \ch{H2} + 1\% \ch{H2O}) & 19958.26 & N & Confidently ruled out. \\
         Clear hybrid atm. (90\% \ch{H2} + 10\% \ch{H2O}) & 1631.44 & N & Confidently ruled out. \\
         Cloudy primary atm. (99\% \ch{H2} + 1\% \ch{H2O}) & 19955.55 & N & Confidently ruled out. \\
         Secondary atm. (100\% \ch{CO2}) & 2.33 & Y & Pure \ch{CO2} atmosphere is possible, but requires $P_{\rm cloudtop} < 1$mbar.\\
         Secondary atm. (100\% \ch{SO2}) & 0.64 & Y & Possible with or without clouds. \\
         Secondary atm. (100\% \ch{H2O}) & 2.84 & Y & Pure \ch{H2O} atmosphere is possible, but requires $P_{\rm cloudtop} < 0.1$mbar. \\
         Secondary atm. (100\% \ch{CH4}) & 2.11 & Y & Pure \ch{CH4} atmosphere is possible, but physically unlikely.\\
         Secondary atm. (\ch{CO}+\ch{CO2}+\ch{SO2}) & 1.66 & Y & Possible with or without clouds. \\
         
         \hline
    \end{tabular}
    \caption{Atmospheric retrievals assuming different compositions applied to the average \matthewred spectrum, for comparison with \protect\cite{luque25}.}
    \label{tab:retrievals}
\end{table*}
\egroup

\section{Discussion}
\label{sec:discussion}

\subsection{Planet Radius}

\begin{figure}
    \centering
    \includegraphics[width=0.45\textwidth]{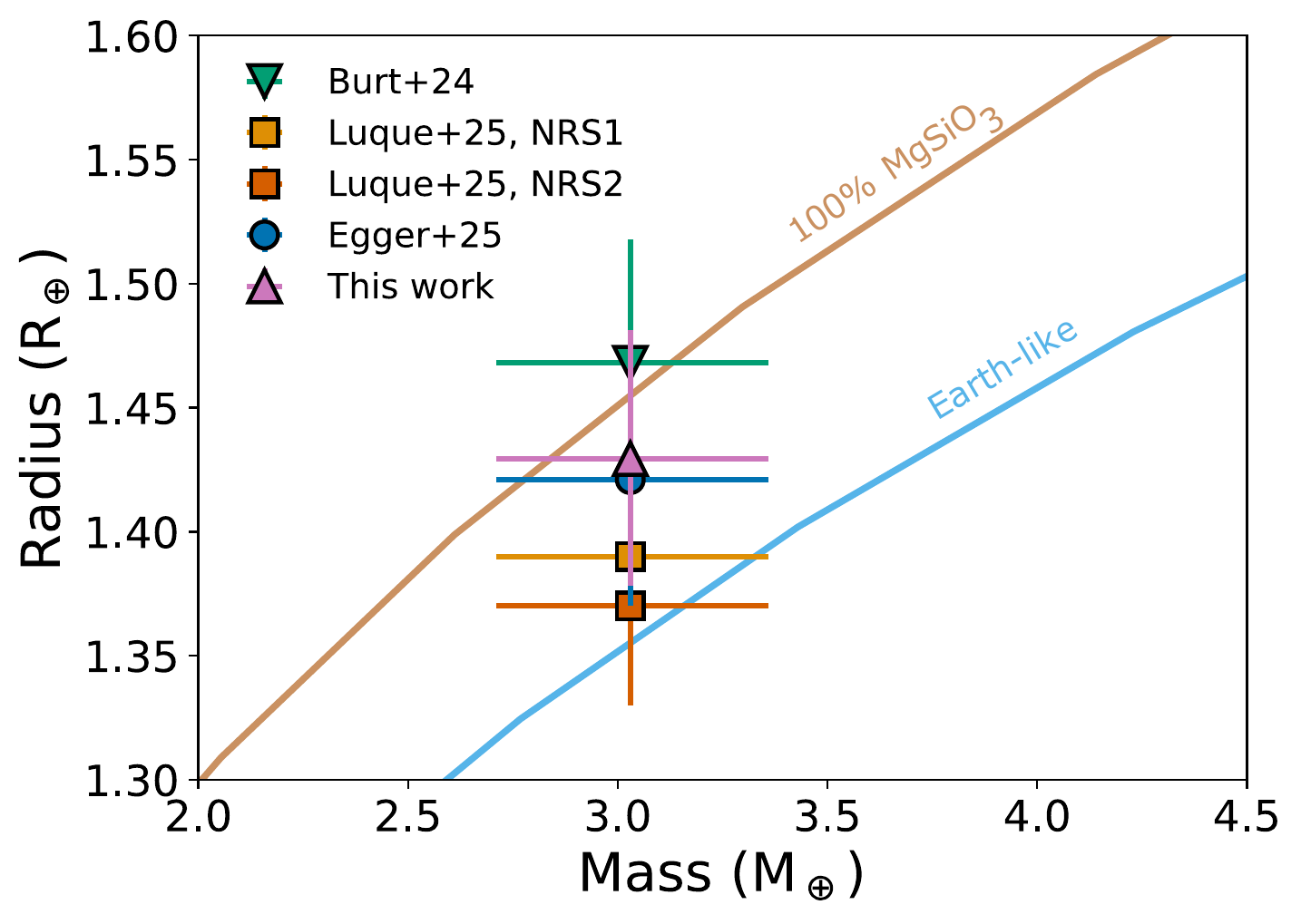}
    \caption{Radius of TOI-1685~b versus the Mass from various recent studies -- Burt+24 \citep{burt24}, Luque+25 \citep{luque25}, Egger+25 \citep{egger25}. The composition curves for a pure silicate and an Earth-like planet are shown in brown and blue, respectively.}
    \label{fig:mass_radius}
\end{figure}

Figure \ref{fig:mass_radius} shows a comparison of the radius values of TOI-1685~b from \cite{burt24}, \cite{luque25}, and \cite{egger25}, compared with the value from our combined white light curve fit from \matthewred. We obtain a value of $1.429^{+0.052}_{-0.051}~R_\oplus$, in good agreement with the fit to the CHEOPS data \cite{egger25}. This places the planet in between the 100\% silicate and Earth-like scenarios. However, due to the degeneracies associated with compositional modelling, the density alone does not constrain the planet's interior \citep[e.g.,][]{dorn15}.

\subsection{Potential Atmosphere}

Although our spectrum is well fit by a flat line, other scenarios, such as high mean molecular weight secondary atmospheres, are not ruled out by the data. We will now discuss the physical plausibility of these atmospheres on TOI-1685~b, as well as the potential for stellar activity in our spectrum, which could make secondary atmospheres harder to detect.

\subsubsection{Secondary Atmospheres}

Our results show that secondary atmospheres are difficult to conclusively rule out on TOI-1685~b, regardless of the composition. These atmospheres would likely be sourced by geochemical outgassing from the planet's interior. In a recent study, \cite{tian24} developed a theoretical framework for the outgassing chemistry of super-Earths and sub-Neptunes, considering both secondary and hybrid atmospheres. They vary the set of input parameters -- the surface pressure, oxidation and sulfidation states of the mantle, and the primordial atmospheric hydrogen, helium, and nitrogen content. They find a rich diversity of possible secondary atmospheres, including some dominated by \ch{H2}. Interestingly, they find that methane-dominated atmospheres are difficult to produce, and require high surface pressures, a reduced mantle, and low magma temperatures. In Section \ref{sec:retrievals}, we find that a pure \ch{CH4} atmosphere could fit the data, but the low evidence combined with results from \cite{tian24}, as well as methane's short photochemical lifetime, makes this an unlikely scenario. Our retrievals also find that the opaque pressure level for pure \ch{H2O} and \ch{CO2} atmospheres is limited to $<0.1-1$ mbar. This could either represent a very high-altitude aerosol, which is physically unlikely, or the planetary surface. Geochemical outgassing is capable of producing almost pure \ch{H2O} or \ch{CO2} atmospheres. However, given the much higher equilibrium temperature of TOI-1685~b (1062 K) compared to Earth (255 K) and Venus (260 K) and assuming higher equilibrium temperatures translate to hotter rocky interiors, the putative stronger volcanism on TOI-1685~b would outgas an atmosphere thicker than 1--100 bar (i.e., those on Earth and Venus). This is unless efficient atmospheric erosion and/or escape could balance the secondary atmosphere at a low pressure level, but this presents a fine-tuning problem \citep[see][for the effect of surface temperatures, which relates to equilibrium temperatures, on outgassed atmosphere pressures]{brachmann25}. Nevertheless, our flat line model in our set of retrievals is favoured by the Bayesian evidence, suggesting there is no evidence for a secondary atmosphere on TOI-1685~b, though it cannot be conclusively ruled out. 

Although we did not investigate these in our modelling, some terrestrial exoplanets could sustain a rock vapor atmosphere through outgassing of a magma ocean. This has been modelled for lava worlds in several studies \citep[e.g.][]{zilinskas22,piette23,seidler24}. Though TOI-1685~b is not a lava world, its maximum dayside temperature is 1390~K, which could result in some melt on the dayside. However, at these temperatures the atmosphere would be extremely thin. Simulations using \texttt{VapoRock} \citep{wolf23} suggest that at this temperature, for an oxygen fugacity of between IW-4 and IW+6, the maximum surface pressure of a vaporised rock atmosphere would be $10^{-5}$~bar. This assumes the entire planet is at 1390~K. At the planet's equilibrium temperature of 1062~K, this drops to $10^{-8}$~bar. Atmospheres with these kinds of surface pressures would not be detectable using low-resolution data, as seen in Section \ref{sec:analysis}. 

Secondary eclipse observations with MIRI's \SI{15}{\micron} filter are a powerful tool to constrain atmospheres on rocky planets \citep[see e.g.,][]{greene23,zieba23,august25,meiervaldes25,fortune25}, and this data could potentially refine our constraints of TOI-1685~b. However, TOI-1685~b is a similar planet to the Hot Rocks target LHS~3780~b, and recent observations show that two eclipses of this planet with MIRI can only rule out \SI{10}{\milli\bar} \ch{CO2} atmospheres ($\sim3\sigma$ confidence) \citep{allen25}. This would not improve on our current constraints for a secondary atmosphere on TOI-1685~b, and improvements would require significant additional time commitment, such as a phase curve.

\subsubsection{Atmospheric Escape}
\label{sec:escape}

TOI-1685 b orbits an early-type M-dwarf (0.45 $M_{\sun}$) under an instellation roughly 200$\times$ that of Earth, exposing it to ionizing irradiation capable of driving catastrophic photoevaporation and placing it beyond the “airless” boundary of the traditional Cosmic Shoreline \citep{zahnle17}. \cite{ji25} explored how the cosmic shoreline varies across hydrodynamic escape models, finding that TOI-1685 b is either marginally or significantly into the regime where a planet born volatile-rich (1 wt\% envelope) loses its atmosphere.  Although a sufficiently massive primordial envelope could allow a hydrogen-dominated atmosphere to persist on TOI-1685 b today, the early boil-off phase limits the initial envelope's size \citep{tang24}, and the observations reported here strongly disfavour a low mean molecular weight envelope.

Alternatively, if efficient atomic line cooling protects metal-rich atmospheres from photoevaporation, as argued by \cite{nakayama22}, then volcanic outgassing could build a secondary atmosphere after primordial loss \citep[e.g., outgassing at a rate of 36 bar/Gyr;][]{kite20b}. However, we calculate with MORS \citep{johnstone21} that the current XUV flux at TOI-1685 b is $\sim 5\times$ greater than the upper limit considered by \cite{nakayama22} of $1000\times$ present Earth levels. Moreover, \cite{chatterjee24} proposed that accounting for ion-electron interactions enables hydrodynamic escape at lower atmospheric temperatures and XUV fluxes compared to \cite{nakayama22}. As yet, there is no consensus on the lifetime of a 1 bar secondary atmosphere on a planet such as TOI-1685 b, owing to the lack of up-to-date simulations of the hydrodynamic escape of metal-dominated super-Earth atmospheres. Despite this uncertainty, and although TOI-1685 is relatively young \citep[$\sim$1 Gyr;][]{egger25}, the extreme irradiation of its innermost planet favours an escaped atmosphere over the presence of a high-altitude cloud layer as the explanation for the flat-line spectrum reported here.

\subsection{Stellar Activity and the Negative Gaussian Feature} 
\label{sec:activity}

TOI-1685 is a relatively inactive M3V star. Stellar activity is inversely correlated with rotation period \citep{skumanich72}. The rotation period of TOI-1685 has been measured by \citet{bluhm21} who found $P_{\rm rot} = 18.66^{+0.71}_{-0.56}$ d, and by \citet{burt24} who found $P_{\rm rot} = 18.2 \pm 0.5$ d. Slow rotation is consistent with the star’s low $v\sin{i}$ \citep{bluhm21}.

\begin{figure}
    \centering
    \includegraphics[width=0.49\textwidth]{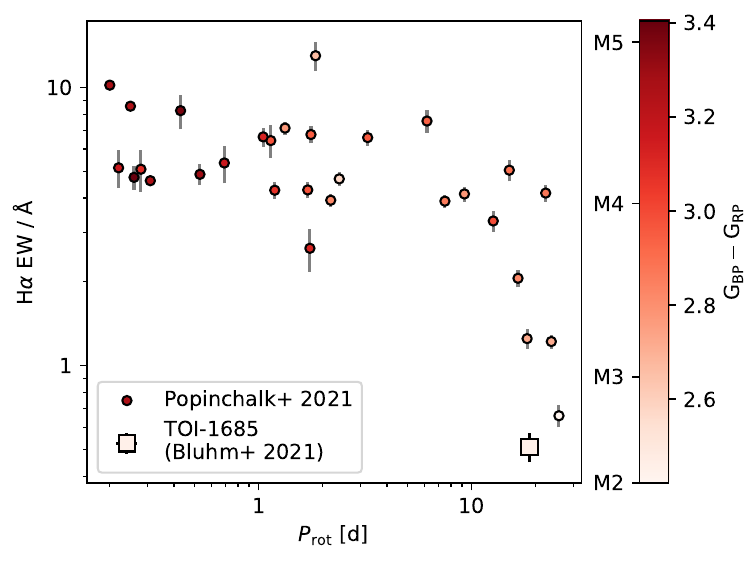}
    \vspace{-5mm}
    \caption{Comparison of the stellar rotation period, photometric color, and H$\alpha$ emission for TOI-1685 \citep[large square,][]{bluhm21} and M2-4 dwarfs from \citep[small circles,][]{popinchalk21}. We expect weak stellar magnetic activity for TOI-1685.}
    \label{fig:activity}
\end{figure}

H$\alpha$ emission is a common proxy for activity in M dwarfs, which correlates with chromospheric heating and magnetic field strength \citep{newton17}. We compare TOI-1685 to M2-4 dwarfs from \citep{popinchalk21} in Figure~\ref{fig:activity} by their rotation periods, photometric colors, and H$\alpha$ equivalent widths (EW) from \citet{bluhm21}. On average, later-type M dwarfs are more active than earlier types, and faster rotators are more active than slower rotators. TOI-1685 is likely to be relatively inactive among stars of similar spectral types.

In Section \ref{sec:gaussian_tests}, we found moderate evidence for a negative Gaussian in one of our transits (29 Feb) in the \matthewred reduction. The existence of this feature is supported by the \tspectro reduction, but is substantially weaker in the \merlinred reduction. The source of this discrepancy is not clear, but did not change with different handlings of limb darkening, spectral binning and Gaussian processes in the light curve fitting.

If the feature is genuine, whilst this could be explained by unusual instrument systematics, recent observations of TRAPPIST-1~e (GO 6456, 9256, PI Allen \& Espinoza) have shown similar features at $\sim$\SI{4.7}{\um} in one of their transits, and high variability at these wavelengths in other transits (Allen et al., \textit{in prep}). Notably, the TRAPPIST-1~e observations use the NIRSpec PRISM instrument, meaning this wavelength region was on a different detector than our observations. This indicates that the feature, if real, is unlikely to be instrumental and may instead be stellar in origin. Although TRAPPIST-1 and TOI-1685 have different expected activity levels, they are both M dwarfs, so could exhibit similar activity signals. Furthermore, some recent work on MHD models of early M dwarfs shows possible stellar activity in this wavelength range due to diatomic molecules like CO \citep{smitha24}. A potentially similar feature has also been seen in a recent observation of GJ 1132 at \SI{4.5}{\micron} \citep{bennett25}, again highlighting the need to better understand M dwarf host stars. Further evidence of stellar contamination in TOI-1685 may be visible in the shorter wavelength NIRISS SOSS observations from JWST GO 4098, once again proving the utility of a large wavelength coverage to confidently characterize small planet transmission spectra, though the separation in time between observations makes connecting the stellar activity more difficult.

The data in this study constitutes one of the best transmission spectra of a super-Earth to-date, and yet we are still unable to conclusively rule out an atmosphere, in part due to sustained uncertainties about stellar activity and instrumental systematics. There are no doubt limitations of JWST, such as the NIRSpec instrumental slope, that we as a community have further work to fully understand. While optical-wavelength data will help address the nature of stellar contamination in particular, improved modeling of M dwarfs is essential for characterizing these host stars and, ultimately, for detecting atmospheres on rocky planets orbiting them.

\section{Conclusions}
\label{sec:conclusions}

In this study, we present the data from our JWST program on TOI-1685~b (GO 4195). We observed four transits of the planet with NIRSpec G395H, and included an additional fifth transit from the phase curve program (GO 3263), already published in \cite{luque25}. We performed three independent data reductions on the five transits, using \matthewred, \tspectro, and \merlinred, obtaining a good agreement across the reductions for each visit. 

We analysed each transit from the \matthewred reduction, and the spectrum averaged over all five transits, using a set of non-physical models to search for evidence of any Gaussian features in the data. Our results show that all five transits can be explained by a flat line, with moderate evidence for a slope in two of the transits. One of the transits shows moderate evidence for a negative Gaussian feature in the spectrum, which we suggest could be due to stellar activity. The overall conclusion of our Gaussian tests is that none of the data show signs of spectral features, indicating either a high molecular weight atmosphere, an atmosphere with high-altitude aerosols, or no atmosphere at all on the planet. 

Using our average transmission spectrum, we then explored the region of parameter space that can be ruled out by our data. For this, we generated a range of atmospheric models, starting with chemical equilibrium, and then moving on to \ch{H2}-molecule mixtures. We varied the opaque pressure level, and either the metallicity or the molecular mixing ratio, and performed a $\chi^2$ analysis to determine the confidence with which a model can be rejected. For chemical equilibrium, we found a minimum metallicity of $\sim1000\times$solar, or a very low opaque pressure level which is unlikely to be physical. This would correspond to a lower limit on the mean molecular weight of $\gtrsim18$, at $\sim5\sigma$. In the \ch{H2}-molecule mixture cases, we found a range of mixing ratio limits depending on the molecule, but a consistent constraint for the mean molecular weight of $\gtrsim10$ at $\sim5\sigma$. This analysis indicates that, even with five transits with JWST, we do not have the sensitivity to probe these heavier, secondary atmospheres that we might expect to be present on rocky exoplanets. 

Finally, we performed a set of simplified atmospheric retrievals on the average spectrum, following the methods of \cite{luque25}, with the goal of determining the gain in sensitivity from five transits over a single one. We tested \ch{H2}-dominated atmospheres with 1\% \ch{H2O} (clear and cloudy cases), a clear \ch{H2}-dominated atmosphere with 10\% \ch{H2O}, cloudy secondary atmospheres with 100\% \ch{CO2}, \ch{SO2}, \ch{H2O}, or \ch{CH4}, and a cloudy mixed secondary atmosphere with \ch{CO2}, \ch{CO}, and \ch{SO2}. Each retrieval had all parameters fixed except the planet radius, the detector offset, and, if relevant, the cloud-top pressure. The exception is the mixed atmosphere case, where the abundances of the molecules were also retrieved. From these retrievals, we can confidently rule out all hydrogen-dominated atmospheres. The cloudy secondary atmospheres cannot be ruled out. However, in the case of pure \ch{H2O} and \ch{CO2}, the retrievals require low cloud-top pressures ($<0.1-1$~mbar), and the pure \ch{CH4} case is physically unlikely. The pure \ch{SO2} case cannot be ruled out, with or without clouds. This analysis represents a significant gain in information over the single transit, which could only conclusively rule out the clear hydrogen-dominated atmospheres \citep{luque25}, but again highlights the challenges with confidently ruling out secondary atmospheres from transmission spectra. 

The results of our study show consistent conclusions with the phase curve analysis, that the planet is well-explained by a bare rock scenario \citep{luque25}. However, it also emphasises some of the limitations of attempting to detect atmospheres on rocky exoplanets with JWST. Due to the heavier nature of the secondary atmospheres we'd expect on these types of planets, observing them in transit is difficult, and conclusively ruling them out remains a challenge. Whether JWST has the precision and senstivity to confidently detect or rule out secondary atmospheres with any number of transits remains to be seen, but extensive programs focused on one or a few targets will aim to finally determine this (e.g., GO 6456, PI Allen \& Espinoza; GO 7073, PI Lustig-Yaeger). Further observations of TOI-1685~b have recently been taken as part of the JWST GO 4098 program, obtaining two transits with NIRISS SOSS. The results from these measurements will hopefully put even tighter constraints on any potential atmosphere (or lack thereof) on TOI-1685~b, as well as provide additional information about stellar activity in the system.

\section*{Acknowledgements}

We thank Can Akin, Prune August, Hannah Diamond-Lowe, Neale Gibson, Mercedes López-Morales, and Anna Lueber for their support with proposal GO 4195.

C.E.F. acknowledges financial support from the European Research Council (ERC) under the European Union’s Horizon 2020 research and innovation program under grant agreement no. 805445. HJH acknowledges support from eSSENCE (grant number eSSENCE@LU 9:3), the Swedish National Research Council (project number 2023-05307), The Crafoord foundation and the Royal Physiographic Society of Lund, through The Fund of the Walter Gyllenberg Foundation. D.K. acknowledges the support from the Swiss National Science Foundation under the grant 200021-231596. E.M.V.
acknowledges financial support from the Swiss National Science Foundation
(SNSF) Postdoc.Mobility Fellowship under grant no. P500PT\_225456/1. RDC acknowledges support from the UK STFC and the AEThER project.

\section*{Data Availability}
Data available at MAST DOI \href{https://doi.org/10.17909/ce7y-8j46}{10.17909/ce7y-8j46}. Data reductions and models available upon request.



\bibliographystyle{mnras}
\bibliography{references} 

\appendix

\section{Detrended light curves}

\begin{figure*}
    \includegraphics[]{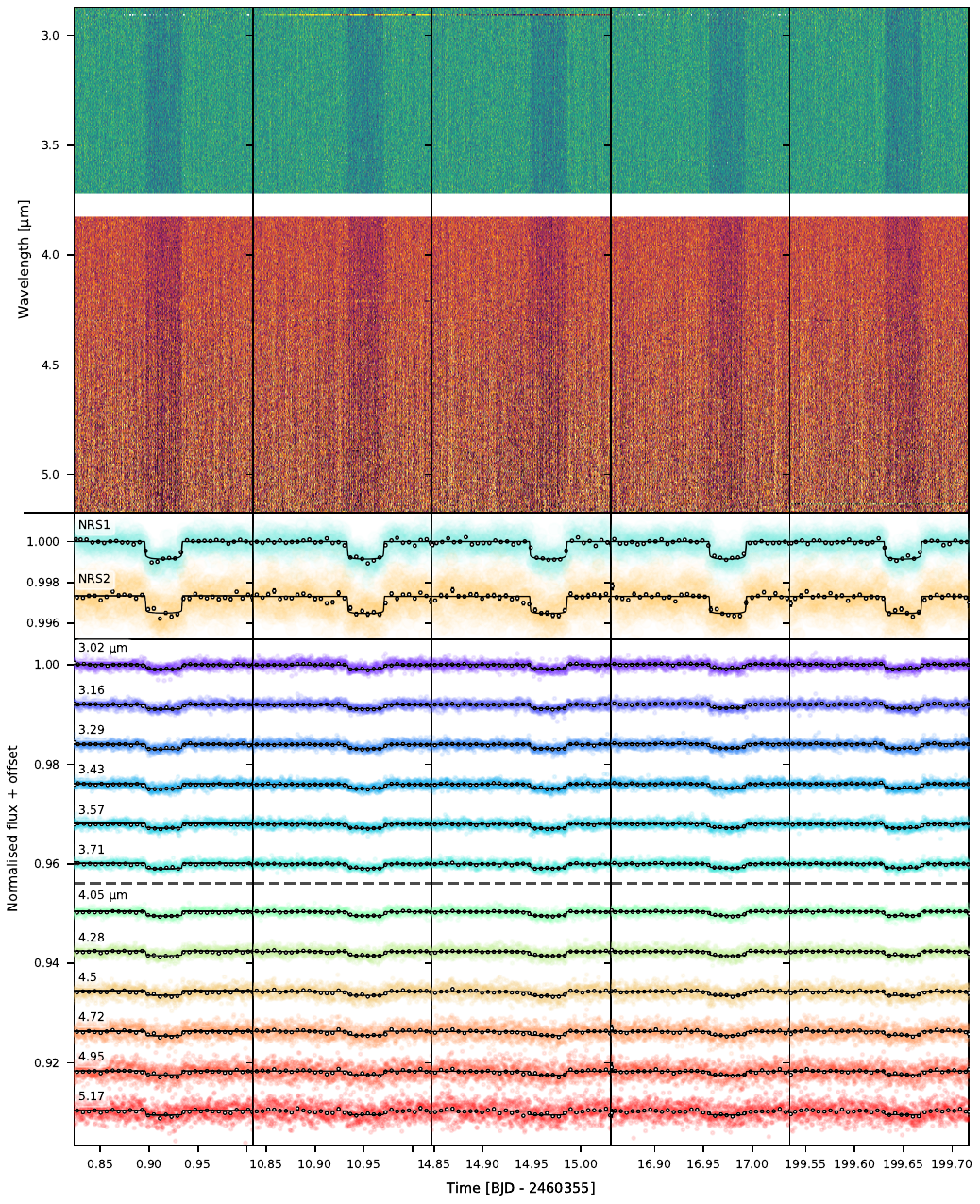}
    \caption{The same light curves and layout as Figure \ref{fig:lcs}, but detrended (divided by the corresponding baseline model in each case).}
    \label{fig:lcs_detrend}
\end{figure*}



\bsp	
\label{lastpage}
\end{document}